\documentclass[letterpaper, journal,12pt, onecolumn]{IEEEtran}
\usepackage[final]{graphicx}
\usepackage{amsmath, amsthm, amssymb,algorithmic}
\usepackage[margin=1in]{geometry}
\usepackage{setspace}
\newtheorem*{mydef}{Corollary}

\newcommand{\lc}{\left\lceil}
\newcommand{\rc}{\right\rceil}
\newcommand{\ben}{\begin{eqnarray}}

\newcommand{\een}{\end{eqnarray}}
\newcommand{\Expect}{\mathbb{E}}

\newtheorem{thm}{Theorem}
\newtheorem*{Corollary}{Corollary}

\newcommand{\trace}{\mathop{\bf tr}}
\newcommand{\vect}{\boldsymbol} %alternative use \vec

\newcommand{\lbb}{\left\{}
\newcommand{\rbb}{\right\}}
\newcommand{\lsb}{\left(}
\newcommand{\rsb}{\right)}
\newcommand{\lmb}{\left[}
\newcommand{\rmb}{\right]}

\title{The Diversity-Multiplexing-Delay Tradeoff in MIMO Multihop Networks with ARQ}

\author{Yao Xie,\thanks{Yao Xie (Email: yaoxie@stanford.edu) is with the
Department of Electrical Engineering at Stanford University.}\quad
\and  Deniz G{\"u}nd{\"u}z,\thanks{Deniz G{\"u}nd{\"u}z  (Email: dgunduz@princeton.edu) was with
the Department of Electrical Engineering, Stanford University, and
with the Department of Electrical Engineering, Princeton
University. He is now with Centre Tecnol\`{o}gic de Telecomunicacions de Catalunya (CTTC), Spain.} \and
\quad Andrea Goldsmith
\thanks{Andrea Goldsmith (Email: andrea@wsl.stanford.edu) is with the Department of
Electrical Engineering at Stanford University.}
\thanks{This work is supported by the Interconnected Focus Center, ONR grant N000140910072P00006, DARPA's ITMANET program, and the Stanford General Yao-Wu Wang Graduate Fellowship.}
} 
\date{\today}

\begin{document}
\maketitle

\begin{abstract}
We study the
tradeoff between reliability, data rate, and delay for half-duplex MIMO multihop
networks that utilize the automatic-retransmission-request (ARQ) protocol both in the asymptotic high signal-to-noise ratio (SNR) regime and in the {f}{i}{n}{i}{t}{e} SNR regime. We propose novel ARQ protocol designs that optimize these tradeoffs. In particular, we
{f}irst derive the diversity-multiplexing-delay tradeoff
(DMDT) in the high SNR regime, where the delay is caused only by
retransmissions. This asymptotic DMDT shows that the
performance of an $N$ node network is limited by the weakest three-node sub-network, and the performance of a three-node sub-network is determined by its weakest link, and, hence, the optimal
ARQ protocol needs to equalize the performance on each link by
allocating ARQ window sizes optimally. This equalization is captured through a novel 
Variable Block-Length (VBL) ARQ protocol that we propose, which achieves
the optimal DMDT. 

We then consider the DMDT in the {f}inite SNR regime, where the delay
is caused by both the ARQ retransmissions and queueing. We characterize the {f}inite SNR DMDT of the {f}ixed ARQ protocol, when an end-to-end delay constraint is imposed, by deriving the probability of message error using an approach that couples the information outage analysis with the queueing network analysis. The exponent of the probability of deadline violation demonstrates that the system performance is again limited by the weakest three-node sub-network.
The queueing  
delay changes the consideration for optimal ARQ design: more
retransmissions reduce decoding error by lowering the information outage probability, but may also increase message drop rate due to delay deadline violations. Hence, the
optimal ARQ should balance link performance while avoiding
signi{f}icant delay. 
We {f}ind the optimal {f}ixed ARQ protocol by solving an optimization problem that minimizes the message error subject to a delay constraint.
\end{abstract}

\vspace{0.3in}
\begin{center}
Submitted to IEEE Trans. Info. Theory, April, 2011.
\end{center}

%\newpage

\section{Introduction}

Multihop relays are widely used for coverage extension in wireless networks when the direct link between the source and destination is weak. The coverage of relay networks can be further enhanced by equipping the source, relays and destination with multiple antennas and using multiple-input-multiple-out (MIMO) techniques for beamforming. Indeed, MIMO can be used either for beamforming, which improves the reliability, or for spatial multiplexing, which increases the data rate \cite{Goldsmith2005}. These dual uses of MIMO gives rise to a diversity-multiplexing tradeoff in point-to-point and multihop MIMO systems, as discussed in more detail below.

Recovery of packets received in error  in multihop networks is usually achieved by automatic retransmission (ARQ) protocols. With an ARQ protocol, on each hop, the receiver feeds back to the transmitter a one-bit
indicator signifying whether the message can be decoded or not. In case
of failure the transmitter retransmits the same
message (or incremental information, e.g., using a Raptor
code \cite{Luby2002}\cite{Shokrollahi2006}) until successful packet reception.
The ARQ protocol can be viewed as either a
one-bit feedback scheme from the receiver to the transmitter, or as
a time diversity scheme employed by the transmitter.
The ARQ protocol improves system reliability at a cost of increased delay. In order to design an effective ARQ protocol for multihop relay networks with MIMO nodes, {f}irst the fundamental tradeoffs between reliability, data rate, and delay of such systems must be determined, and then the protocol performance can be compared to this theoretical performance limit. 

A fundamental tradeoff in designing point-to-point MIMO systems is the tradeoff
between reliability and data rate, characterized by the diversity-multiplexing tradeoff (DMT). The asymptotic DMT was introduced in
\cite{ZhengTse03diversity} focusing on the asymptotically high SNR regime. The {f}inite SNR DMT was presented in \cite{Narasimhan2006}. The DMT
has also been used to characterize the performance of classical three-node relay networks, with a direct link between the source and the destination, when the nodes have single-antenna (SISO) or multiple antennas for various relaying strategies \cite{LanemanTseWornell2003}, \cite{YukselErkip2007},
\cite{GunduzKhojastepourGoldsmithDMTMIMO2008}.  The DMTs for the
amplify-and-forward (AF) and decode-and-forward (DF) relaying strategies are
discussed in \cite{LanemanTseWornell2003}. Several extensions of the amplify-and-forward strategy have been proposed recently, including the rotate-and-forward relaying \cite{PedarsaniLevequeYang2010} and {f}lip-and-forward relaying \cite{PedarsaniLevequeYang2010FF} strategies, which employ a sequence of forwarding matrices to create an arti{f}icial time-varying channel within a single slow fading transmission block in order to achieve a higher diversity gain. 
A dynamic
decode-and-forward (DDF) protocol, in which the relay listens to the
source transmission until it can decode the message and then
transmits jointly with the source, is proposed in
\cite{AzarianElGamal2005} and its DMT performance is shown to dominate the {f}ixed AF and DF schemes. The DDF protocol is
shown to achieve the optimal DMT performance in MIMO multihop relay
networks in \cite{GunduzKhojastepourGoldsmithDMTMIMO2008}. In this paper, we restrict our attention to multihop networks using the DF relaying strategy, since it enables us design an optimal ARQ protocol for MIMO multihop relay networks, as we will show later.

%Multiple input multiple output (MIMO) systems can provide
%increased data rates through multiplexing by creating multiple
%parallel channels, or they can provide robustness against channel
%variations by increasing diversity. Relaying is yet another technique to increase the diversity of
%wireless systems by exploiting the resources of nearby terminals. The destination may combine the signals from the source and the relay
%terminals in decoding the underlying message to obtain the cooperative diversity provided by relaying \cite{Sendonaris:CT:03,LanemanTseWornell2003}. In another setting, the multihop relaying, \hl{where relays are used for coverage extension}, each terminal receives the signal only
%from the previous terminal in the route. Multihop
%relaying is widely used in practice to increase coverage or to
%improve energy efficiency

Here we consider the diversity-multiplexing-delay tradeoff (DMDT), which was introduced in \cite{GamalDamen06themimo}  as an extension
of the DMT to include the delay dimension. Here the notion of delay is the time from the arrival
of a message at the transmitter until the message is successfully decoded at the receiver, also known as the ``sojourn time'' in
queueing systems. Delays are incurred for two reasons: (1)
ARQ retransmissions: messages are retransmitted over each hop until correctly decoded at the corresponding receiver, and (2) queueing delay: ARQ results in a queue of messages to be retransmitted at the transmitter. Most works on DMDT assume in{f}inite SNR for the
asymptotic analysis, and the queueing delay has been largely
neglected. This is because in the high SNR regime, retransmission is
a rare event \cite{HollidayGoldsmithPoor2008}. With this asymptotic
in{f}inite SNR assumption, \cite{GamalDamen06themimo} presents the
DMDT for a point-to-point MIMO system with ARQ, \cite{Tabet:IT:07} studies
the DMDT for cooperative relay networks with ARQ and single-antenna
nodes, and \cite{AzarianElGamalSchniter2008} proves the
DMDT-optimality of ARQ-DDF for the multiple access relay and the
cooperative vector multiple access channels with single antenna
nodes. However, the asymptotically high SNR regime does not capture the operating conditions of typical wireless systems in practice, where errors during transmission
attempts are not rare events \cite{HollidayGoldsmithPoor2008}. Hence, to fully characterize the DMDT performance in the {f}inite SNR regime, we must bring the queuing delay into the problem formulation.  For a point-to-point MIMO system with a delay constraint and no feedback link, the tradeoff  between the error caused by outage due to insuf{f}icient code length, and the error caused by delay exceeding a given deadline, has been studied in \cite{KittipiyakulEliaJavidi2009} using large deviation analysis.

One of the goals of our paper is to study the effects of dynamic ARQ on the DMDT in relay networks. Hence, we consider a line network 
in which a node's transmission is only received by adjacent nodes in the line. This is a reasonable approximation for environments where received power falls off sharply with distance (i.e., the path loss exponent is large).
For this multi-hop channel model we show that the optimal ARQ protocol requires dynamic allocation of the ARQ transmission rounds based on the instantaneous channel state, and we obtain its exact DMDT characterization. The more general case where non-adjacent nodes receive a given node's transmission is signi{f}icantly more complicated, and the optimal DMT is unknown for this case even with a single relay \cite{Tabet:IT:07}.

The contribution of this paper is two-fold: (1) we characterize the DMDT of multihop MIMO relay networks in both the asymptotically high SNR regime and in the {f}inite SNR regime where, in the latter, queuing delay is incorporated into the analysis;  (2) we design the optimal ARQ protocol in both regimes. Our work extends the DMDT analysis of a point-to-point MIMO system presented in \cite{HollidayGoldsmithPoor2008} to MIMO multihop relay networks. In the {f}irst part of the paper, we derive the DMDT in the asymptotic high SNR regime, where the delay is caused by retransmissions only. For a certain
multiplexing gain, the diversity gain is found by studying the information
outage probability. An information outage occurs when the receiver fails to
decode the message within the maximum number of retransmission
rounds allowed. Based on this formulation, for some multihop relay networks a closed-form expression for the DMDT can be identi{f}ied, whereas for general multihop networks, determining the DMDT can be cast as an optimization problem that can be solved numerically. The DMDT of a general multi-hop network can be studied by decomposing the network into three-node sub-networks. Each three-node sub-network consists of any three neighboring nodes in the network and the corresponding links between them. The asymptotic DMDT result
shows that the performance of the multihop MIMO network, i.e., its DMDT, is determined by the three-node sub-network with the minimum DMDT. The DMDT of the three-node sub-network is again determined by its weakest link. 
Hence, the optimal ARQ protocol should balance the
link DMDT performances on each hop by allocating ARQ window sizes among the hops. From
this insight, we present an adaptive variable block-length (VBL)
ARQ protocol and prove its DMDT optimality.

Next, we study the DMDT in the {f}inite SNR regime, in which the delay is caused by both retransmissions and queueing. We introduce an end-to-end
delay constraint such that a message is dropped once its delay exceeds this
constraint. We characterize the  {f}inite SNR DMDT by studying the probability of message error, which is dominated by two causes: the information outage event and the missing deadline event, when the block length is su{f}{f}iciently long \cite{ZhengTse03diversity}. Our approach couples the information-theoretical outage probability \cite{Narasimhan2006} with queueing network analysis. In contrast to the analysis under asymptotically high SNR, this does not yield closed-form DMDT expressions; however, it leads to a practically more relevant ARQ protocol design. The end-to-end delay that takes the queueing delay into consideration introduces one more factor into the DMDT tradeoff and the associated optimal ARQ protocol design. Speci{f}ically, allocating more transmission rounds to a link may improve
its diversity gain and, hence, lower the information outage probability; however,
it also increases the queueing delay and, hence, may also increase the overall error probability as more messages are
dropped due to the violation of the deadline. Thus, an optimal ARQ protocol in the
{f}inite SNR regime should balance these con{f}{l}{i}cting goals: our results will show that this leads to equalizing the DMDT performance of the links. 

We formulate the optimal ARQ protocol
design as an optimization problem that minimizes the probability of message error under a given delay constraint. The end-to-end delay constraint requires us to take into
account the message burstiness and queueing delays, which are known
to be the main obstacles in merging the information-theoretical physical
layer results with the network layer analysis
\cite{EphremidesHajek1998}. We bridge this gap by modeling the MIMO
multihop relay network as a queueing network. However, unlike in traditional queueing network theory, e.g., \cite{BolchGreinerdeMeer2006, BertsekasGallager1992}, the multihop network with half-duplex relay nodes is not a standard queue tandem, because node $i$ along the multihop queue tandem must wait to complete reception of the previous message by the node $i+2$ before it can transmit to node $i+1$ in the tandem. Another difference between our analysis and traditional queueing theory is that we study the amount of time a message waits in the queue (similar to
\cite{HollidayGoldsmithPoor2008}) 
%(In replying to Deniz's question: waiting time is also studied in queue theory? The answer is: yes. But the theory for waiting for is much less than that for the number of packets waiting in the queue. The waiting time is basically a reflected random walk for a single server station. When we have a queue-tandem - multiple service station in series, the distribution of end-to-end service time is not clear and often only its tail distribution was studied. Here we have half-duplex relays, which adds further constraint - the related end-to-end delay is a new problem and hasn't been considered. Also as explained in the following couple of sentences.), 
rather than just the number of messages awaiting transmission. This poses a
challenge because the distribution associated with these random delays is hard to obtain
\cite{Harrison1973}, unlike the distribution of the number of
messages for which a product form solution is available
\cite{BolchGreinerdeMeer2006}. In \cite{BisnikAbouzeid2006} delay is studied by using a closed queue model and diffusion approximation. We derive the exponent of the deadline missing probability in our half-duplex multihop MIMO network by adapting the large deviation argument used in \cite{Ganesh1998}. The expression of the exponent again demonstrates that the system performance (in terms of the exponent) of multi-hop network with half-duplex relays is determined by the three-node sub-network with the minimum exponent.

The remainder of this paper is organized as follows. Section
\ref{sec:model} introduces the system model and the ARQ protocol.
Section \ref{sec:DMDT_asymptotic} presents the asymptotic DMDT
analysis for various ARQ protocols while proving the DMDT optimality of the VBL ARQ. Section
\ref{sec:DMDT_finite_SNR} presents the {f}inite SNR DMDT with queueing delays, including some illustrative examples. {F}inally, Section \ref{sec:conclusion} concludes the paper and discusses some future directions.

\section{System Model and ARQ Protocols} \label{sec:model}

\subsection{Channel Model}

\begin{figure}[h]
\centering \includegraphics[width=5in]{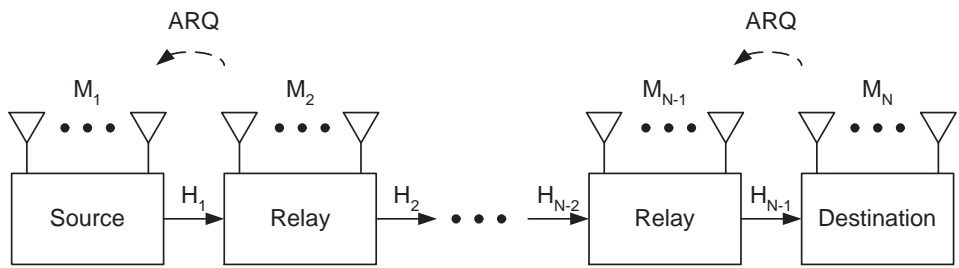} \caption{MIMO
multihop relay network with ARQ. The number of antennas on the
$i$th node is $M_i$, and $\vect{H}\in \mathcal{C}^{M_{i+1}\times M_i}$ is the channel matrix from node $i$ to node  $(i+1)$. } \label{Fig:scheme_all}
\end{figure}

Consider an $N$-node multihop MIMO relay network. Node 1 is the
source, node $N$ is the destination, while nodes 2 through $N-1$
serve as relays. Node $i$ has $M_i$ antennas for $i = 1, \cdots, N.$
The system model is illustrated in {F}ig. \ref{Fig:scheme_all}. We
denote this MIMO relay network as $(M_1, M_2, \cdots, M_N)$. At the source, the message is encoded by a space-time encoder
and mapped into a sequence of $L$ matrices, $\{\vect{X}_{1,l} \in
\mathcal{C}^{M_1\times T}: l = 1, \cdots, L\}$, where $T$ is the
block length, i.e., the number of channel uses of each block, and $L$ is the maximum number of end-to-end total ARQ rounds that can be used to transmit each message from the
source to the destination. The
rate of the space-time code is $R$.

We de{f}ine one ARQ round as the transmitter sending a whole block code
of the message to the receiver.  We assume that the relays use the DF
protocol: node $i$, $2\leq i \leq N-1$, decodes the message, and
reencodes it with a space-time encoder  into a sequence of $L$
matrices $\{ \vect{X}_{i,l} \in \mathcal{C}^{M_i \times T}: l = 1,
\cdots, L \}$.
%(since the ARQ transmission can take up to $L$ rounds).
The channel between node $i$ and node ($i+1$) is given by:
\begin{eqnarray}
\vect{Y}_{i,l} = \sqrt{\frac{SNR}{M_i}} \vect{H}_{i,l}
\vect{X}_{i,l} + \vect{W}_{i,l}, \quad 1 \leq l \leq L,
\end{eqnarray}
where  $\vect{Y}_{i,l} \in \mathcal{C}^{M_{i+1} \times
T}$, $i = 1, \cdots, N-1$, is the received signal at node $(i+1)$
in the $l$th ARQ round. Channels are assumed to be frequency
non-selective, block Rayleigh fading and independent of each
other, i.e., the entries of the channel matrices $\vect{H}_{i,l}
\in \mathcal{C}^{M_{i+1}\times M_i}$ are independent and
identically distributed (i.i.d.) complex Gaussian with zero mean
and unit variance. The additive noise terms $\vect{W}_{i,l}$ are
also i.i.d. circularly symmetric complex Gaussian with zero mean and unit variance. The
forward communication links and ARQ feedback links only exist between
neighboring nodes.

Other assumptions we have made for the channel model are as follows:
\begin{itemize}
 
\item[(i)] We consider half-duplex relays, that is, the relays cannot transmit and receive at the same time. 

\item[(ii)] We assume a short-term power constraint at each node
for each block code, given by $
\Expect\{\trace(\vect{X}_{i,l}^\dag \vect{X}_{i,l})\} \leq M_i $,
$\forall i, l$.
Here $\Expect\{\cdot\}$ denotes expectation, and $\dag$ denotes
the Hermitian transpose. A long-term power constraint would allow us to
adapt the transmit power and achieve power control gain, as we
brie{f}ly discuss later in the paper. In the following results we assume a short-term power
constraint in order to focus on the diversity gain achieved by the
ARQ protocol.

\item[(iii)] We consider both the long-term static channel model, in which
$\vect{H}_{i,l} = \vect{H}_i$ for all $l$, i.e. the channel state
remains constant during all the ARQ rounds for a hop and is
independent from hop to hop; and the short-term static channel model,
where $\vect{H}_{i,l}$ are i.i.d. but not identical for the same
$l$.
The long-term static channel assumption is the worst-case in terms
of the achievable diversity with a maximum of $L$ ARQ rounds
\cite{GamalDamen06themimo}, because there is no time diversity
gain. The long-term static channel model may be
suitable for modeling indoor applications such as Wi-{F}i, while the short-term static channel model suits applications with higher mobility, such as outdoor cellular systems. 
\end{itemize}

\subsection{Multihop ARQ Protocols}

Consider a family of multihop ARQ protocols, in which the following
standard ARQ protocol is used over each hop. The receiver in each hop tries to decode the message after or during
one round, depending on whether the synchronization is per-block based or
per-channel-use based. Once it is able to decode the message, a
one bit acknowledgement (ACK) is fed back to the
transmitter that triggers the transmission of the next message.
After one ARQ round, if the receiver cannot decode the message, a
negative acknowledgement (NACK) is fed back to the
transmitter. Then the transmitter sends the next block of the code that
carries additional information for the same message. The retransmission over the $i$th hop continues for a maximum
number of  $L_i$ rounds, called the ARQ window size. Once the ARQ
window size is reached without successful decoding of the message, the message is discarded, causing an
information outage. Then the next message is transmitted. The sum of the ARQ window sizes is upper
bounded by $L>0$, where
\begin{equation}
\sum_{i=1}^{N-1} L_i\leq L.
\end{equation}
We consider several
ARQ protocols with different ways to allocate the available ARQ windows among different hops:
\begin{itemize}
\item[(i)] A {f}ixed ARQ protocol, which allocates a {f}ixed ARQ window
size of $L_i$ for the transmitter of node $i$, $i = 1, \cdots, N-1$ such that $\sum_{i=1}^{N-1}L_i = L$.

\item[(ii)] An adaptive ARQ protocol, in which the allocation of the ARQ
window size per hop is not {f}ixed but adapted to the channel
state. The transmitter of a node can keep retransmitting as long
as the total ARQ window size of $L$ has not been reached. We further consider two types of adaptive ARQs based on different synchronization levels:

\begin{itemize}
\item[(1)] {F}ixed-Block-Length (FBL) ARQ protocol: The
synchronization is per-block based. The transmission of a message over each hop spans an integer number of ARQ rounds.

\item[(2)] Variable-Block-Length (VBL) ARQ protocol: The
synchronization is per-channel-use based. The receiver can send an ACK as soon as it can decode the message, and
the transmitter starts transmitting a new message without waiting until the beginning of the next channel block. VBL has a {f}iner time resolution than FBL and is more ef{f}icient  in using the available channel block, at a cost of higher synchronization complexity.
\end{itemize}

\end{itemize}

We assume that the ARQ feedback links has zero-delay and no error.

\section{Asymptotic DMDT}\label{sec:DMDT_asymptotic}

We characterize the tradeoff among the data
rate (measured by the multiplexing gain $r$), the
reliability (measured by the diversity gain $d$), and the
delay by the asymptotic DMDT of a system with ARQ. Following the framework of
\cite{ZhengTse03diversity} and \cite{GamalDamen06themimo}, we assume that the rate of transmission depends on the operating SNR
$\rho$, and consider a family of space time codes with block rate $R(\rho)$ scaling with
the logarithm of SNR as 
\begin{equation}
R(\rho) = r\log\rho.
\end{equation}

\subsection{Diversity Gain}

In the high SNR regime, the diversity gain is de{f}ined as the SNR exponent of the message error probability \cite{ZhengTse03diversity}. It is shown in
\cite{ZhengTse03diversity} that the message error probability $P_e(\rho)$ is dominated by the information outage probability $P_{out}(\rho)$ whenÊ the block-length is suf{f}iciently large. In the following we make this assumption.  The \textit{information outage event} is the event that the
accumulated mutual information at the receiver within the allowed ARQ window size does
not meet the data rate of the message and, therefore the receiver
cannot decode the message.  Hence, the diversity gain for a family of codes is de{f}ined as:
\begin{eqnarray}
d(r) \triangleq -\lim_{\rho
\rightarrow \infty} \frac{\log P_{e}(\rho)}{\log \rho}.\label{def_diversity}
\end{eqnarray}

The DMT of an $M_1\times M_2$ MIMO system is denoted by $d^{(M_1, M_2)}(r)$ and de{f}ined as the supremum of the diversity gain $d(r)$ over all families of codes. DMT of a point-to-point MIMO system is characterized in
\cite{ZhengTse03diversity} by the following theorem:
\begin{thm}\label{DMT}
For a suf{f}iciently long block-length, theÊ
DMT $d^{(M_1,M_2)}(r)$ is given by
the piece-wise linear function connecting the points $(r, (M_1 -
r)(M_2 - r)),$ for $r = 0, \cdots, \min(M_1, M_2)$.
\end{thm}

%In the following,
%the asymptotic DMDT is expressed in terms of the multiplexing gain
%$r$, to simplify notation. The actual DMDT expression in terms of the effective multiplexing gain can be obtained by using (\ref{r_e}). 

\subsection{Asymptotic DMDT}\label{sec:asymptoticDMDT}

To characterize the asymptotic DMDT for a multihop network in the high SNR regime, we need the following quantity. Assume that the channel inputs at both the source and the
relays are Gaussian with identity covariance matrices. De{f}ine $M_i^* = \min\{M_i, M_{i+1}\}$, for $i = 1, \cdots, N-1$. For the
long-term static channel, let $\lambda_{i,1}, \cdots,
\lambda_{i,M_i^*}$ be the nonzero eigenvalues of $\vect{H}_i
\vect{H}_i^\dag$, for $i = 1, \cdots, N-1$. Suppose $\lambda_{i,j} =
\rho^{-\alpha_{i,j}}$, for $j = 1, \cdots, M_i^*$, $i = 1, \cdots, N-1$. At
high SNR, we can approximate the channel capacities
${C}_i(\vect{H}_i)\triangleq \log \det \left(\vect{I} +
\frac{\rho}{M_i} \vect{H}_i \vect{H}_i^\dag\right)$ as
$
C_i(\vect{H}_i) \doteq \log \rho^{S_i(\vect{\alpha}_i)}$  \footnote{Here the exponential equality $\doteq$ is de{f}ined as $f(\rho)
\doteq \rho^c$, if $\lim_{\rho\rightarrow \infty}\frac{\log
f(\rho)}{\log \rho} = c$. The exponential inequalities
$\dot{\leq}$ and $\dot{\geq}$ are de{f}ined similarly. },
where
\begin{equation}
S_i(\vect{\alpha}_i) \triangleq \sum_{j = 1}^{M_i^*}(1-\alpha_{i,j})^+,\label{S_def}
\end{equation}
$(x)^+ \triangleq \max\{x, 0\}$,
and the vector $\vect{\alpha}_i \triangleq [\alpha_{i,1}\cdots \alpha_{i,M_i^*}].$
This $S_i(\vect{\alpha}_i) $ plays an important role in the asymptotic DMDT analysis. The closer the SNR exponents $\alpha_{i,j}$'s are to unity, the closer
the channel matrix is to being singular. Similarly, we can de{f}ine $\{\alpha_{i,
j}^l\}$ in the short-term static channel model and the corresponding matrix $\vect{\alpha}_i \in
\mathbb{R}^{M_i^*\times L}$ as $[\vect{\alpha}_i]_{k,l} \triangleq
{\alpha}_{i,k}^l$.

Proofs for the asymptotic DMDT analysis rely on the notion of \textit{decoding time}, which is the time at which the
accumulated information reaches $R(\rho)$. In the case of the
short-term static channel, for the FBL ARQ and other block-based ARQ
protocols, the decoding time for the $i$th node is given by
\begin{eqnarray}
t_i \triangleq \inf \lbb t \in \mathbb{Z}^+: \sum_{l = 1}^{t}
C_i(\vect{H}_{i,l})\geq r\log \rho \rbb  \doteq  \inf\lbb t  \in
\mathbb{Z}^+: \sum_{l = 1}^{t} S_i(\vect{\alpha}_i^l) \geq r \rbb,
\label{stop_3}
\end{eqnarray}
where $\mathbb{Z}^+$ denotes the set of  positive integers. For the VBL ARQ and
other non-block-based ARQ protocols, the decoding time is given by
\begin{eqnarray}
t_i  \triangleq  \inf\lbb t \in \mathbb{R}: \sum_{l = 1}^{\lfloor t
\rfloor} S_i(\vect{\alpha}_i^l) + (t - \lfloor t \rfloor)
S_i(\vect{\alpha}_i^{\lfloor t \rfloor+1}) \geq r \rbb, \label{stop_2}
\end{eqnarray}
where $\lfloor x\rfloor$ denotes the largest integer smaller than
$x$. Similarly we can de{f}ine the decoding time for the long-term
static channel model. We can view the accumulated mutual information as a
random walk with random increments $S_i(\vect{\alpha}_i^l)
> 0$ and stopping boundary $r$.

In the following, we {f}irst state our results for the three-node network $(M_1, M_2, M_3)$, and then extend them to the general $N$-node network.

\subsubsection{Long-Term Static Channel}

The DMDT of the {f}ixed ARQ protocol in the case of the long-term static channel
is given by the following theorem:
\begin{thm}\label{thm_static_ARQ}
\textit{With the long-term static channel assumption, the DMDT of the {f}ixed ARQ protocol for a three-node MIMO multihop network with window sizes $L_1$ and $L_2$, $L_i \in \mathbb{Z}^+$, $L_1 + L_2 \leq L$, is given by:}
\begin{eqnarray}
d_F^{(M_1, M_2, M_3)}(r,L_1, L_{2}|L_1+L_2\leq L) = \min_{i = 1, 2} \lbb d^{(M_i,
M_{i+1})}\lsb \frac{r}{L_i}\rsb\rbb. \label{dmt_multi_ARQ}
\end{eqnarray}
\end{thm}
%\vspace{0.3in}
%
\noindent Proof: See Appendix \ref{app:thm_static_ARQ}. 

Consistent with our intuition, (\ref{dmt_multi_ARQ}) shows that
the performance of a three-node network is limited by the weakest link. This
implies that if there were no constraint for the $L_i$'s to be
integers, the optimal choice should equalize the
diversity-multiplexing tradeoff of all the links, i.e.,
\begin{eqnarray} d^{(M_1, M_2)}\lsb
\frac{r}{L_1}\rsb = d^{(M_2, M_3)}\lsb \frac{r}{L_2}\rsb. \label{link_perf_match}
\end{eqnarray}
With the integer constraint we choose the integer $L_i$'s such that the minimum of $d^{(M_i, M_{i+1})}\left(\frac{r}{L_i}\right)$ for $i = 1, 2$ is maximized. 

The DMDT of the FBL ARQ protocol is a piece-wise linear function
characterized by the following theorem:
\begin{thm}\label{thm_block_DDF}
\textit{With the long-term static channel assumption, the DMDT of
the FBL ARQ protocol for a three-node MIMO multihop network is given
by}
\begin{eqnarray}
d_{FBL}^{(M_1, M_2, M_3)}(r, L) = \min_{l_i \in \mathbb{Z}^+ :  l_1 + l_2
= L-1} \lbb d^{(M_1, M_2)}\lsb
\frac{r}{l_1}\rsb
+ d^{(M_2, M_3)}\lsb
\frac{r}{l_2}\rsb\rbb.
\end{eqnarray}
\end{thm}
%\noindent where we define $f_{M_i, M_{i+1}}(\infty) \triangleq 0$.
%\vspace{0.3in}

\noindent Proof: See Appendix \ref{app:thm_block_DDF}. 

%The minimization in the theorem statement again captures the fact that the performance of a \hl{three-node} network is limited by the weakest link.

The DMDT of the VBL ARQ protocol cannot always be expressed in closed-form, but can be written as the solution of an optimization problem, as stated in the following theorem.
\begin{thm}\label{thm_DDF}
\textit{With the long-term static channel assumption, the DMDT of the VBL ARQ protocol for a three-node MIMO multihop network is given by
\begin{eqnarray}
d_{VBL}^{(M_1, M_2, M_3)}(r, L) = \inf_{(\vect{\alpha}_1, \vect{\alpha}_2) \in \mathcal{O}} h(\lbb
\alpha_{i,j} \rbb), \label{ddf1}
\end{eqnarray}
where $ h(\lbb \alpha_{i,j} \rbb)\triangleq \sum_{i=1}^{2}
\sum_{j = 1}^{M_i^*} (2j - 1 + |M_i - M_{i+1}|)\alpha_{i,j}$. The set $\mathcal{O}$ is de{f}ined as}
\begin{equation}
\begin{split}
\mathcal{O} \triangleq & \left\{(\vect{\alpha}_1, \vect{\alpha}_2)\in
\mathbb{R}^{M_1^*}\times \mathbb{R}^{M_2^*}: \right. \\
& \alpha_{i,1}\geq \cdots\geq \alpha_{i,M_i^*}\geq 0,  i = 1, 2,
\left.
\frac{S_1(\vect{\alpha}_1)S_2(\vect{\alpha}_2)}{S_1(\vect{\alpha}_1) + S_2(\vect{\alpha}_2)}
 < \frac{r}{L}
\rbb,
\end{split} \label{feas_VBL2}
\end{equation}
and this is the optimal DMDT for a three-node network in the long-term static channel.
\end{thm}
%
%\vspace{0.3in}

\noindent Proof: See Appendix \ref{app:thm_DDF}.

Note that the DMDT of the VBL ARQ protocol in the
three-node network, under the long-term static channel assumption,
is similar to the DMT of DDF without ARQ given in
\cite{GunduzKhojastepourGoldsmithDMTMIMO2008}, with proper scaling
of the multiplexing gain. The optimization problem in (\ref{ddf1}) can be shown to be convex using techniques similar to Theorem 3 in \cite{GunduzKhojastepourGoldsmithDMTMIMO2008}.

We have closed-form solutions for some speci{f}ic cases where the
optimization problem in (\ref{ddf1}) has a simple form and can be
solved analytically. For example, for a $(M_1, 1, M_3)$
network (\ref{ddf1}) becomes
\begin{eqnarray}
d_{VBL}^{(M_1, 1, M_3)}(r, L) = \inf_{\alpha_{1,1}, \alpha_{2,1}} &&M_1\alpha_{1,1} + M_3 \alpha_{2,1} \nonumber\\
\mbox{subject to}&&
\frac{(1-\alpha_{1,1})^+ (1-\alpha_{2,1})^+}{(1-\alpha_{1,1})^+ + (1-\alpha_{2,1})^+} < \frac{r}{L}, \nonumber \\
&& \alpha_{i,1} \geq 0, \quad i = 1, 2.
\end{eqnarray}
 The DMDT for this case (and two
other special cases) is given by the following corollary:
\begin{mydef}
With the long-term static channel assumption the DMDT of the VBL ARQ protocol
\textit{\begin{itemize} \item[(1)]
for a $(M_1, 1, M_3)$  MIMO multihop network is given
by
\begin{eqnarray}
d_{VBL}^{(M_1, 1, M_3)}(r,L) = \left\{%
\begin{array}{ll}
    \min\{M_1,M_3\}\frac{1-2r/L}{1-r/L}, & 0 \leq r\leq L/2; \\
    0, & \hbox{otherwise.}
\end{array}%
\right.
\end{eqnarray}
\item[(2)] for a $(1,M,1)$ 
MIMO multihop network is given by 
\begin{eqnarray}
d_{VBL}^{(1, M, 1)}(r, L) = \left\{
\begin{array}{ll}
M\frac{1-2r/L}{1-r/L}, & 0\leq r \leq L/2;\\
0, & \hbox{otherwise.}
\end{array}
\right.
\end{eqnarray}
\item[(3)] for a (2, 2, 2) 
MIMO multihop network is given by
\begin{eqnarray}
d_{VBL}^{(2, 2, 2)}(r,L) = \left\{%
\begin{array}{ll}
    \frac{2(4-5r/L)}{2-r/L}, & {0 \leq r \leq L/2;} \\
    \frac{3-4r/L}{1-r/L}, & {L/2 \leq r \leq 2L/3 ;} \\
    \frac{4(1-r/L)}{2-r/L}, & {2L/3  \leq r \leq L;} \\
    0, & \hbox{otherwise}.
\end{array}%
\right. \nonumber
\end{eqnarray}
\end{itemize}}
\end{mydef}

%\vspace{0.3in}

\subsubsection{Short-Term Static Channel}

The DMDT of the {f}ixed and the FBL ARQ under the short-term static
channel assumption are similar to those under the long-term static
channel assumption, with additional scaling factors for DMDTs of each hop due to the time diversity gain. 

\begin{thm}\label{thm_st}
\textit{With the short-term static channel assumption, the DMDT of
the FBL ARQ protocol for a three-node MIMO multihop network is given by
\begin{eqnarray}
d_{FBL}^{(M_1, M_2, M_3)}(r, L) = \min_{l_i \in \mathbb{Z}^+ : l_1 + l_2
= L - 1} \lbb \sum_{i = 1}^{2} l_i d^{(M_i, M_{i+1})}\lsb
\frac{r}{l_i}\rsb\rbb.\nonumber
\end{eqnarray}}
\end{thm}
%
%\vspace{0.3in}
\noindent Proof: See Appendix \ref{app:thm_st}.

\begin{thm}\label{thm_st_fractional}
With the short-term static channel assumption, the DMDT of
the VBL ARQ protocol for a three-node  MIMO multihop network is given by
\begin{eqnarray}
d_{VBL}^{(M_1, M_2, M_3)}(r, L) = \inf_{(\vect{\alpha}_1, \vect{\alpha}_{2}) \in
\mathcal{G}} \tilde{h}\left(\lbb \alpha_{i,j}^l \rbb\right),
%\label{ddf}
\end{eqnarray}
where $\tilde{h}\left(\lbb \alpha_{i,j}^l \rbb\right)\triangleq
\sum_{i=1}^{2} \sum_{j = 1}^{M_i^*} \sum_{l = 1}^{L}\lsb 2j - 1
+ |M_i - M_{i+1}|\rsb \alpha_{i,j}^l$, and the set $\mathcal{G}$ is de{f}ined as
\begin{eqnarray}
\mathcal{G} &\triangleq& \lbb
(\vect{\alpha_1}, \vect{\alpha}_{2})\in\mathbb{R}^{M_1^* \times
L}\times \mathbb{R}^{M_{2}^*\times L}: \right.
 \nonumber\\
 &&~~~\left.  \alpha_{i,
1}^l\geq \cdots \geq \alpha_{i, M_i^*}^l \geq 0, \forall i,
l, \quad t_1 + t_2 > L
 \rbb.\label{VBL_region}
\end{eqnarray}
The $t_i$'s, de{f}ined in (\ref{stop_2}), depend on the $\vect{\alpha}_i$'s. This is the optimal DMDT for a three-node MIMO multihop network in the short-term static channel.
\end{thm}
\noindent Proof: See Appendix \ref{app:thm_st_fractional}.

\subsection{DMDT of an $N$-Node Network and Optimality of VBL}

Next, we extend our DMDT results to general $N$-node MIMO multihop networks. Note that in our model, since each transmitted signal is received only by the next node in the network, the transmission over the $i$th hop does not interfere with other transmissions.  We will show the DMDT of this more general network is a minimization of the DMDTs of all its three-node sub-networks, due to half-duplexing and multihop diversity. The multihop diversity \cite{GunduzKhojastepourGoldsmithDMTMIMO2008} captures the fact that we allow simultaneous transmissions of multiple node pairs in half-duplex relay networks. For example, while node $i$ is transmitting to node $(i+1)$, node $(i+2)$ can also transmit to node $(i+3)$. This effect allows us to split a message into pieces, which are transmitted simultaneously in the network to increase the multiplexing gain. Using this rate-splitting scheme, we can prove the DMDT optimality of the VBL ARQ protocol. Due to their {f}ixed block length, we are only able to provide upper and lower bounds for DMDTs of {f}ixed ARQ and FBL ARQ in an $N$-node network.

\begin{thm}\label{thm_n_VBL}
With the long-term or short-term static channel assumption, the DMDT of the VBL ARQ for an $N$-node MIMO multihop network is given by
\begin{equation}
d_{VBL}^{(M_1, \cdots, M_N)}(r, L) = \min_{i = 1, \cdots, N-2} d_{VBL}^{(M_i, M_{i+1}, M_{i+2})}(r, L),
\end{equation}
and this is the optimal DMDT for an $N$-node network.
\end{thm}
\begin{proof}
See Appendix \ref{app:thm_n_VBL}.
\end{proof}
Theorem \ref{thm_n_VBL} says that the DMDT of an $N$-node system is determined by  the smallest DMDT of its three-node sub-networks. The minimization in Theorem \ref{thm_n_VBL} is over all possible three-node sub-networks instead of pairs of nodes, due to the half-duplex constraint: each low-rate piece of message has to wait for the previous piece to go through two hops before it can be transmitted. Theorem \ref{thm_n_VBL} also says that the VBL ARQ is the optimal ARQ protocol in the general multi-hop network.

\begin{thm}\label{thm_n_fixed}
With the long-term or short-term static channel assumption, the DMDT of {f}ixed ARQ for an $N$-node network is lower bounded and upper bounded, respectively, by
\begin{equation}
d_{F}^{(M_1, \cdots, M_N)}(r, L_1, \cdots, L_{N-1}) \geq \min_{i = 1, \cdots, N-2} d_{F}^{(M_i, M_{i+1}, M_{i+2})}\left(\frac{L_{\max}}{L}r, L_i, L_{i+1}\;\middle\vert\;L_i + L_{i+1} \leq L_{\max}\right),
\end{equation}
and
\begin{equation}
d_{F}^{(M_1, \cdots, M_N)}(r, L_1, \cdots, L_{N-1}) \leq \min_{i = 1, \cdots, N-2} d_{VBL}^{(M_i, M_{i+1}, M_{i+2})}(r, L), 
\end{equation}  
where $L_{\max} \triangleq \max_{i=1}^{N-2}\left\{L_i + L_{i+1}\right\}$.
\end{thm}
\begin{proof}
See Appendix \ref{app:thm_n_fixed}.
\end{proof}
%\textbf{Remark:} The lower and upper bounds for the fixed ARQ protocol may meet in some special cases, which determines the exact DMDT of the fixed ARQ protocol. For example, consider a multihop network $(2, 2, 2, 2)$, and $L_i = 2$, for $i = 1, 2, 3$. In this case, $L_{\max} = 4$. The DMDT scheme

\begin{thm}\label{thm_n_FBL}
With the long-term or short-term static channel assumption, the DMDT of the FBL ARQ for an $N$-node network is lower bounded and upper bounded, respectively, by
\begin{equation}
d_{FBL}^{(M_1, \cdots, M_N)}(r, L) \geq \min_{i = 1, \cdots, N-2} d_{VBL}^{(M_i, M_{i+1}, M_{i+2})}(r,  L - N),\label{FBL_lower}
\end{equation}
and
\begin{equation}
d_{FBL}^{(M_1, \cdots, M_N)}(r, L) \leq \min_{i = 1, \cdots, N-2} d_{VBL}^{(M_i, M_{i+1}, M_{i+2})}(r, L).
\label{FBL_upper}
\end{equation} 
\end{thm}
\begin{proof}
See Appendix \ref{app:thm_n_FBL}.
\end{proof}

\begin{figure}[h] \centering
\includegraphics[width=3.5in]{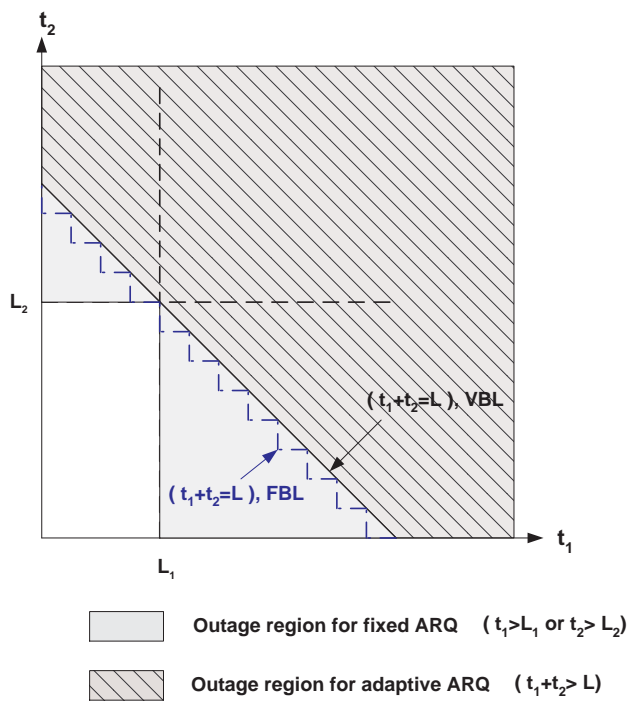} \caption{Outage regions of the {f}ixed ARQ, and two adaptive ARQs: 
the FBL and the VBL ARQs, with $L_1 + L_2 = L$.}
\label{Fig:tau}
\end{figure}

An intuitive explanation for the DMDT optimality of the VBL ARQ is as follows.  Recall that $t_i$ is the number of channel blocks, including retransmissions, needed to decode the message over the $i$th hop. For a three-node network, we can illustrate the information outage region in the region
of $t_1\times t_2$ values  as in {F}ig. \ref{Fig:tau}.
The outage region of the VBL ARQ is smaller than those of the {f}ixed and the FBL ARQ. Due to its per-block based synchronization, the outage region boundary of the FBL ARQ is a piecewise approximation to that of the VBL ARQ. In the high SNR regime, we formalize the above intuition in the following corollary to Theorem \ref{thm_n_FBL}.

\begin{Corollary}
With the long-term or short-term static channel assumption, for an $N$-node MIMO multihop network, the DMDT of the FBL ARQ converges to that of the VBL  ARQ when $L\rightarrow\infty$.
\label{thm_conv}
\end{Corollary}
%\vspace{0.3in}
\begin{proof}
Using (\ref{FBL_lower}) and (\ref{FBL_upper}), when $L\rightarrow\infty$, 
\begin{equation}
\begin{split}
\min_{i = 1, \cdots, N-2} d_{VBL}^{(M_i, M_{i+1}, M_{i+2})}(r, L)
 \geq d_{FBL}^{(M_1, \cdots, M_N)}(r, L)
& \geq  \min_{i = 1, \cdots, N-2} d_{VBL}^{(M_i, M_{i+1}, M_{i+2})}(r,  L[1- N/L])\\
& \xrightarrow{L\rightarrow \infty} \min_{i = 1, \cdots, N-2} d_{VBL}^{(M_i, M_{i+1}, M_{i+2})}(r,  L).
\end{split}\label{FBL_near}
\end{equation} \end{proof}

\subsection{Power Control Gain with Long Term Power Constraint}

With the long-term power constraint and channel state information at the transmitter (CSIT), we can employ a power control strategy to further improve diversity. Let the SNR in the $l$th round be $\rho(l) = \rho^{g(l)}$, where $\rho$ is the average SNR, and $g(l)$ is the function de{f}ining the power control strategy. In the high SNR regime, similar to (\ref{S_def}) we can approximate channel capacities as %\begin{eqnarray}
$C_i (\vect{H}_i) \doteq \log \rho^{S'_i(\vect{\alpha}_i)}$, where
$S'_i(\vect{\alpha}_i) = \sum_{j = 1}^{M_i^*}\lsb g(l) -
\alpha_{i,j}\rsb^+$. Hence, with power control, all the asymptotic DMDT results in the
previous sections hold with $S_i(\vect{\alpha}_i)$ replaced by
$S'_i(\vect{\alpha}_i)$. 

\subsection{Examples for Asymptotic DMDT}

\begin{figure}[h]
\centering \includegraphics[width=3.1in]{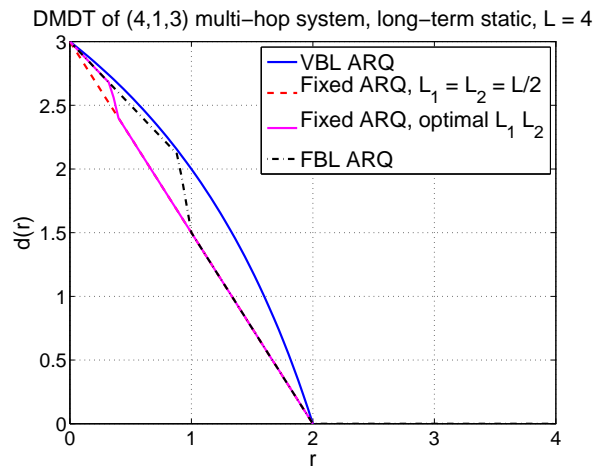}  \caption{The DMDT for a
(4,1,3) multihop network with $L = 4$.} \label{Fig:dmt_314}
\end{figure}

\begin{figure}[h]\centering
\includegraphics[width=3.2in]{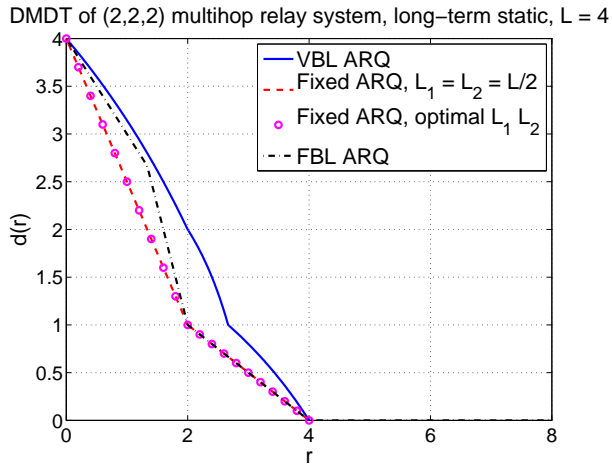} \caption{The DMDT for a (2,2,2) multihop network with $L = 4$.} \label{Fig:dmt_222}
\end{figure}

In this section we show some illustrative examples for the asymptotic DMDT. We {f}irst consider the long-term static channel model. For a three-node $(4,1,3)$
multihop network with $L=4$, {F}ig. \ref{Fig:dmt_314} shows
the DMDT of the {f}ixed ARQ with $L_1 = L_2 = L/2$, of
the per-hop-performance-equalizing $L_1$ and $L_2$
satisfying (\ref{link_perf_match}), % in Corollary \ref{col3},
as well as the DMDTs of the FBL and the VBL ARQs.
%Fig. \ref{Fig:block_413} shows the DMDT for FBL
%ARQ (the red line) as a minimum of two lines (different
%exponents), as defined in Theorem \ref{thm_block_DDF}.
Note that the DMDT of the VBL ARQ in {F}ig. \ref{Fig:dmt_314} is the optimal DMDT for the (4, 1, 3) network. We also consider a (2, 2, 2) network, whose DMDTs are shown in {F}ig. \ref{Fig:dmt_222}. 

\begin{figure}
\centering
\includegraphics[width=3in]{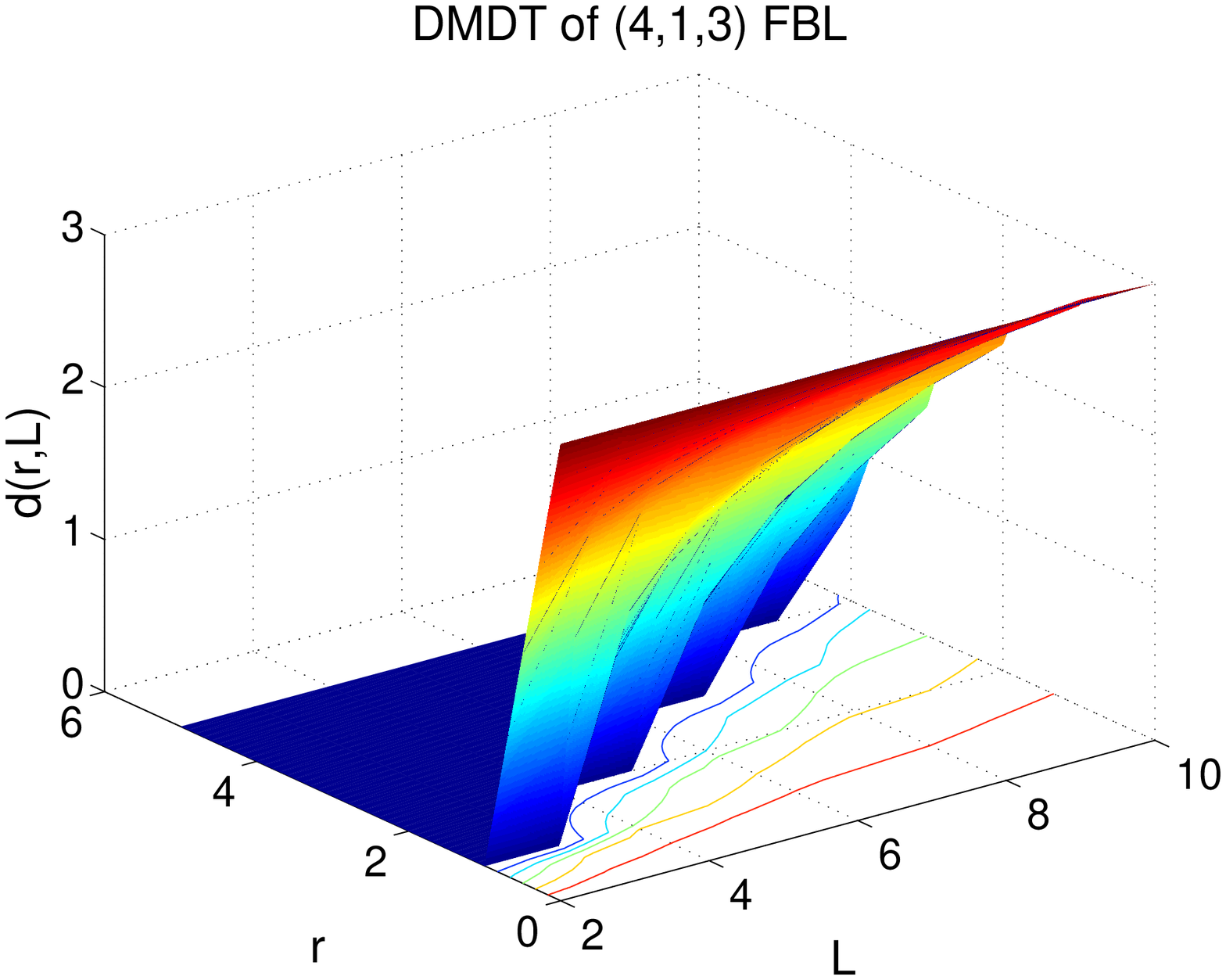}
\includegraphics[width=3in]{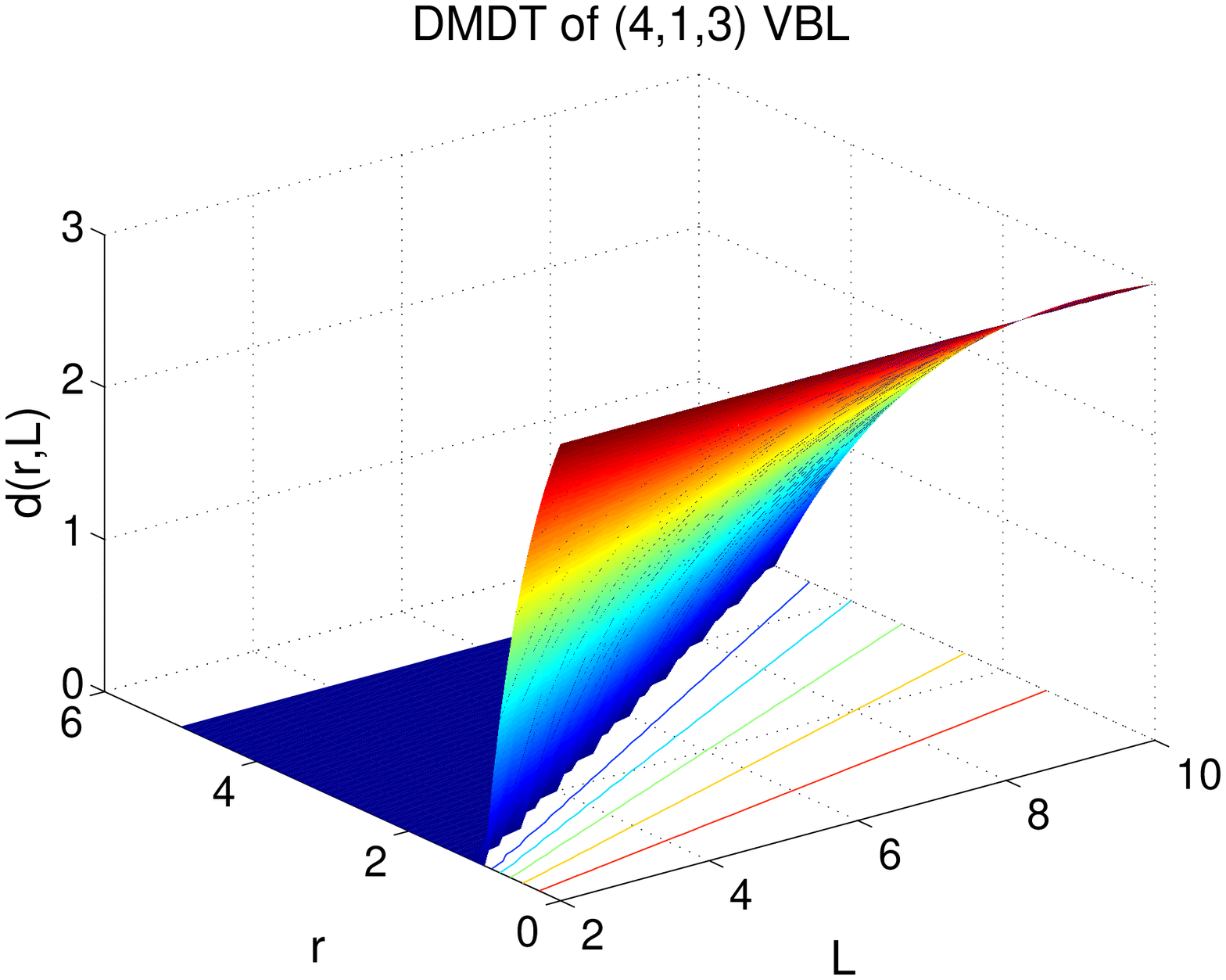}
 \caption{The three-dimensional DMDT surface
for a (4,1,3) network, with the FBL ARQ (left) and the VBL
ARQ (right).} \label{Fig:dmt_413_L}
\end{figure}

\begin{figure}
\centering \includegraphics[width=3in]{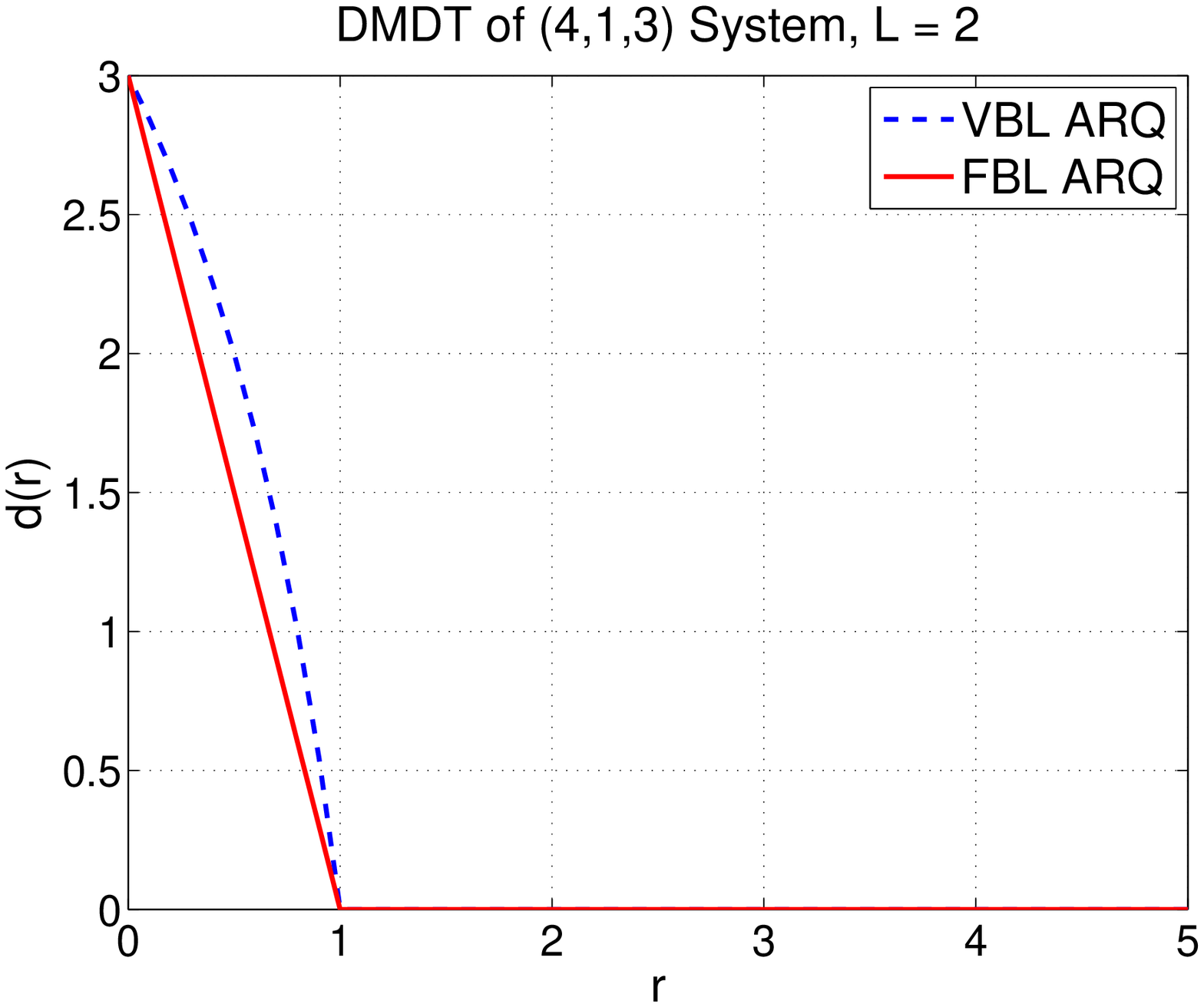}
\includegraphics[width=3in]{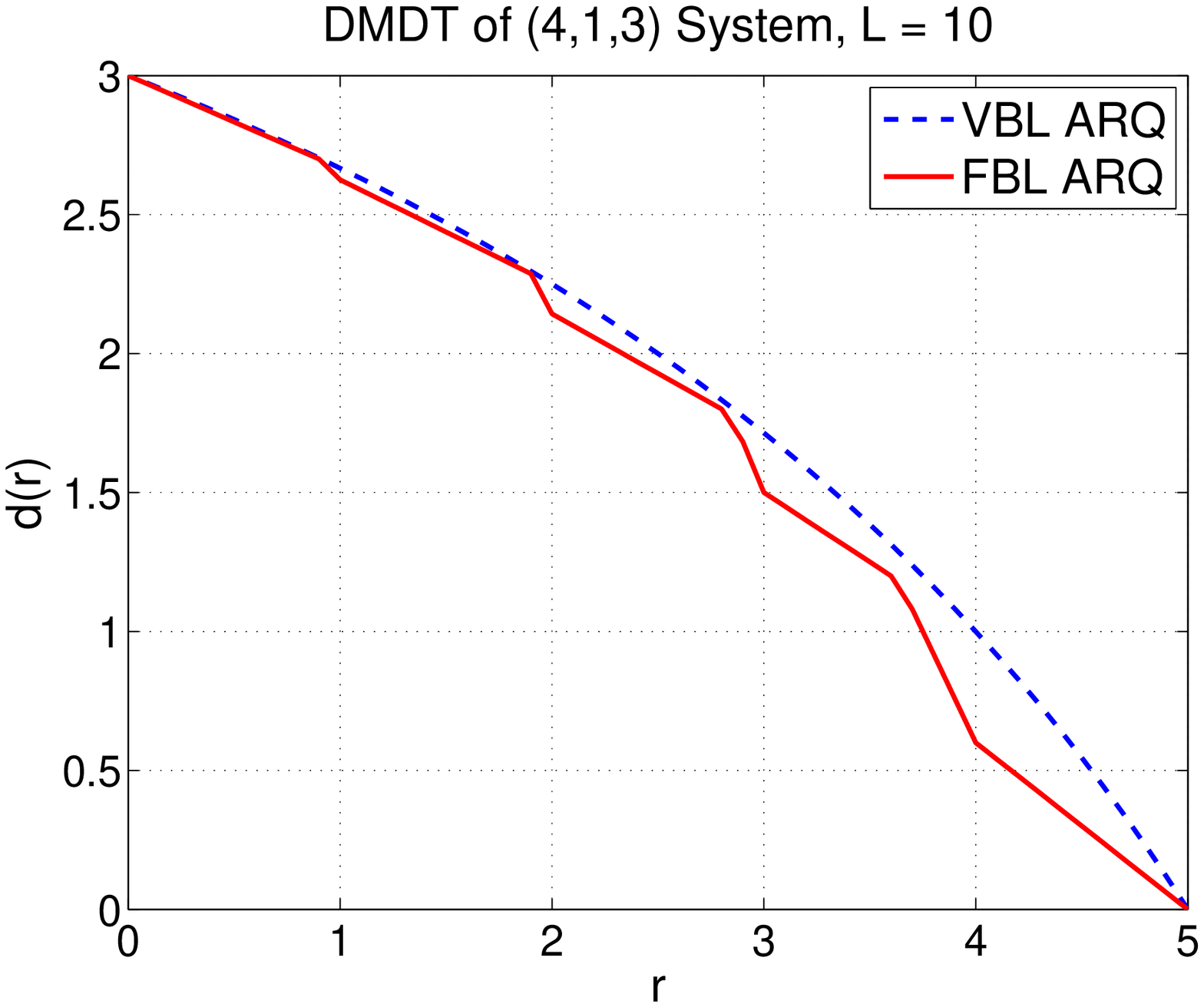} \caption{The slices of the DMDT surface in {F}igure \ref{Fig:dmt_413_L} at $L = 2$ (left)
and at $L = 10$ (right).} \label{Fig:dmt_413_L_slice}
\end{figure}

{F}ig. \ref{Fig:dmt_413_L} presents the three-dimensional DMDT surface
of the VBL and the FBL ARQs, respectively, for the $(4,1,3)$ multihop network. Note that as $L$ increases, the diversity gain at a given $r$ increases for both the FBL and the VBL ARQ protocols. Also note that the DMDT surface of the FBL ARQ is piecewise and that of the VBL ARQ is smooth due to their different synchronization levels. {F}ig.
\ref{Fig:dmt_413_L_slice} illustrates the cross sections of the surfaces
in {F}ig. \ref{Fig:dmt_413_L} at $L = 2$ and $L = 10$, which demonstrates
the convergence of the DMDTs proved in Theorem \ref{thm_conv}.

\begin{figure}
\centering
\includegraphics[width=3in]{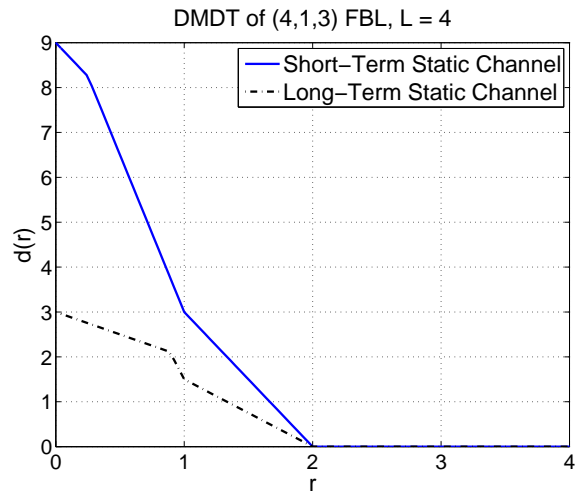}
\caption{The DMDT for a (4, 1, 3) multihop network in the long-term static
channel versus that in the short-term static channel.}
\label{Fig:dmt_413_s_l}
\end{figure}

Next we consider the short-term static channel model. The DMDT of the (4,1,3) multihop network using the FBL ARQ is shown in {F}ig. \ref{Fig:dmt_413_s_l}. Note that the asymptotic DMDT in the short-term static channel model is not
necessarily a multiple $L$ of the corresponding DMDT in the long-term static
channel model, which differs from the point-to-point MIMO channel
\cite{GamalDamen06themimo}, where the asymptotic DMDT in the short-term static channel model is a multiple $L$ of the corresponding DMDT in the long-term static channel model.

\section{{F}inite SNR DMDT With Delay
Constraint}\label{sec:DMDT_finite_SNR}
Ê
In analyzing the {f}inite SNR DMDT, we add a practical end-to-end delay constraint: each message has to reach the destination before the deadline; otherwise it is discarded. We characterize the {f}inite SNR DMDT by studying the probability of message error. With the delay constraint brought into the picture, the probability of message error has two components: the \textit{information outage probability} and the \textit{deadline missing probability}, which we analyze using the {f}inite SNR DMT introduced in \cite{Narasimhan2006} and the queueing network analysis, respectively. In the {f}inite SNR regime, the multiplexing gain is de{f}ined as:
\begin{eqnarray}
r \triangleq \frac{R}{\log_2(1+M_r \rho)},
\end{eqnarray}
where $M_r$ is the number of antennas at the receiver. In the following we only consider the long-term static channel model and the {f}ixed ARQ protocol. We {f}irst introduce the queueing network model. 

\subsection{Queueing Network Model}

The messages enter the network at the source node and exit from the destination node, forming an open queue. Messages arrive at
the source node as a Poisson process with a mean message inter-arrival time of 
$\lambda$ blocks. As in the previous sections, the unit of time is one block of the channel consisting of $T$ channel uses. The end-to-end delay constraint is $k$ blocks. Each node can be viewed as a service station
transmitting (possibly with several retransmissions) a message to
the next node. The time node $i$ spends to successfully transmit a message to node $(i+1)$ is called the service
time of the $i$th node, which depends on the channel
state and is upper bounded by the ARQ window size $L_i$.  The allocated ARQ window sizes satisfy $\sum_{i} L_i\leq k$.
%The allocated ARQ window size $L_i$ satisfies $L_i\leq k$ (but their sum is not necessarily upper-bounded by $L$) (\hl{To Deniz: I think we probably don't need the constraint $\sum_i L_i = L \leq k$, because it is taken care of by the optimization problem already. The solution $\sum_i L_i = L > k$ will have greater cost function value than the other case) As you could observe from the numerical example, this never becomes a solution}. 

As an approximation, we assume that the random service
time at node $i$ for each message is i.i.d. with an exponential
distribution of mean $\mu(L_i)$ (the actual service time has value  distributed in the interval of $[0, L_i]$). Here $\mu(L_i)$, which we derive later, is the actual average service time of the ARQ process when the ARQ window size is $L_i$.  With these assumptions we can treat
each node as an $M/M/1$ queue. This approximation makes the
problem tractable and characterizes the qualitative behavior of
MIMO multihop networks. The messages enter the buffer and are processed based on the {f}irst-come-{f}irst-served (FCFS) rule. We assume $\mu(L_i) + \mu(L_{i+1})< \lambda$, $i = 1, \cdots, N-2$, so the queues are stable, i.e., the waiting time at
a node does not grow unbounded.

Burke's theorem (see \cite{BolchGreinerdeMeer2006}) says
that in an $M/M/1$ queue with Poisson arrivals, the messages leave the server as a Poisson process. Hence messages arrive at each relay (and the destination node) as a Poisson process with rate $(1-p_i)/\lambda$,  where $p_i$ is the probability that a message is dropped.
% due to information outage or deadline missing in the $i$th hop. 
When the SNR is reasonably high, we can assume the message drop probability is small and hence $1 - p_i \approx 1$.

\subsection{Probability of Message Error}

Denote the total queueing delay experienced by the $n$th message transmitted from the source to the destination as $W_n$, and the number of transmissions  needed by node $i$ to transmit the $n$th message to node $(i+1)$ as $t_n^i$, if the transmitter can use any number of rounds. For the $n$th message, if it is not discarded due to information outage, the total ``service time''  is $S_n = \sum_{i=1}^{N-1} \min\{t_n^i, L_i\}$ and the random end-to-end delay is 
$
D_n = W_n + S_n. 
$
Recall for the {f}ixed ARQ, the message is dropped once the number of retransmissions exceeds the ARQ window size of any hop, or the end-to-end delay exceeds the deadline. Hence, the message error probability of the $n$th message 
can be written as
\begin{eqnarray}
P_e &=& P 
\lbb \cup_{i=1}^{N-1}\{t_n^i>L_i\}\rbb+ P\lbb
\cap_{i=1}^{N-1}\{t_n^i \leq L_i\} 
\cap\{D_n> k\} 
\rbb. \label{eqn22}
\end{eqnarray}
The {f}irst term in (\ref{eqn22}) is the message outage probability:
\begin{equation}
P_{out}(\{L_i\}|\rho) \triangleq  P 
\lbb \cup_{i=1}^{N-1}\{t_n^i>L_i\}\rbb,\label{out_finiteSNR}
\end{equation} 
which is identical for any message $n$ since channels are i.i.d. The second term in (\ref{eqn22}) is related to the deadline missing probability, and can be rewritten as
\begin{equation}
\begin{split}
P\lbb \cap_{i=1}^{N-1}\{t_n^i \leq L_i\} 
\cap\{D_n> k\}   \rbb = [1-P_{out}(\{L_i\}|\rho) ] P\lbb D_n > k \middle\vert \cap_{i=1}^{N-1}\left\{ t_n^i \leq L_i \right\} \rbb
\end{split}.
\end{equation}
De{f}ine the stationary deadline missing probability:
\begin{equation}
P_{deadline}\lsb \{L_i\}|\rho, k\rsb\triangleq  \lim_{n\rightarrow\infty}P\lbb D_n > k \middle\vert \cap_{i=1}^{N-1}\left\{ t_n^i \leq L_i \right\} \rbb.\label{deadline_finiteSNR}
\end{equation}
In the following we derive  (\ref{out_finiteSNR}) and (\ref{deadline_finiteSNR}).

%%
%The probability of message error
%depends on the ARQ window size allocation $L_i$, delay constraint $k$, multiplexing rate $r$ and SNR $\rho$. 
%In the following we study these two probabilities separately. 

\subsubsection{Information Outage Probability}

Since channels in different hops are independent, (\ref{out_finiteSNR})  becomes
%(similar to what we had in the asymptotic regime.)
\begin{equation}
\begin{split}
P_{out}\lsb\{L_i\}|\rho\rsb &= \sum_{i=1}^{N-1} P\lbb t_n^i> L_i\rbb 
= \sum_{i=1}^{N-1}P\lbb L_i C_i(\vect{H}_i) < r\log_2(1+M_{i+1}\rho)\rbb \\
&\triangleq \sum_{i=1}^{N-1}P_{out,i}\lsb L_i|\rho \rsb,
\end{split}
\label{finite_SNR_pout}
\end{equation}
which is a sum of the per-hop outage probabilities $P_{out, i}\lsb L_i|\rho \rsb$.
%i.e., the probability that the accumulated information for $i$th link up to the transmission of the $L_i$ round is still below the required level $r\log_2(1+M_{i+1}\rho)$ for a successful transmission. 
Using results in \cite{Narasimhan2006} for point-to-point MIMO, we have 
\begin{equation}
P_{\rm{out}, i}\lsb L_i|\rho \rsb =
  \sup_{(b_1, \cdots b_{M_i^*}) \in \mathcal{B}_i} \prod_{l=1}^{M_i^*} \frac{\gamma \lsb M_{i+1} - M_i + 2l - 1, \frac{M_i}{\rho}[(1+M_{i+1} \rho)^{b_l} - (1+ M_{i+1} \rho)^{b_{l-1}}] \rsb}{\Gamma(M_{i+1} - M_i + 2l - 1)},\label{per-hop-out}
\end{equation}
where the set $\mathcal{B}_i$ is given by
\begin{equation}
\mathcal{B}_i = \lbb (b_1, \cdots, b_{M_i^*}) \left|
b_{l-1} <  b_l < \frac{\frac{r}{L_i}-\sum_{k=1}^{l-1}b_k}{M_i^* - l +1}
\right., l = 1, \cdots M_i^*-1; b_{M_i^*} = \frac{r}{L_i} - \sum_{k=1}^{M_i^*-1}b_k \rbb, 
\end{equation}
$\gamma(m, x) \triangleq \int_{0}^x t^{m-1} e^{-t} dt$ is the incomplete gamma function, and $\Gamma(m) \triangleq (m-1)!$ for a positive integer $m$. 
%The optimization problem in (\ref{per-hop-out}) does not have a closed-form solution in general. However, 
For orthogonal space-time block coding (OSTBC), we can derive a closed-form $P_{out,i}(L_i|\rho)$ using techniques similar to \cite{Narasimhan2006}:
\begin{equation}
P_{\rm out, OSTBC}\lsb\{L_i\}|\rho\rsb = \sum_{i=1}^{N-1} P\lbb r_s L_i\log_2\lsb 1+\frac{\rho}{M_{i}}\|\vect{H}_i\|_F^2 \rsb\leq
r\log_2(1+ M_{i+1}\rho)\rbb ,\label{32}
\end{equation}
where $\|\vect{A}\|$ denotes the Frobenius norm of a matrix $\vect{A}$, and the spatial code rate $r_s$ is equal to the average number of independent constellation symbols transmitted per channel use. For example, $r_s = 1$ for the Alamouti space-time code \cite{Alamouti1998}.  When $\vect{H}_i$ is Rayleigh distributed, its Frobenius norm has the Gamma(1, $M_i \cdot M_{i+1}$) distribution. Hence, (\ref{32}) becomes:
\begin{equation}
P_{\rm out, OSTBC}(\{L_i\}|\rho) = \sum_{i=1}^{N-1}\frac{1}{(M_i\cdot M_{i+1}-1)!}\gamma\left(\frac{M_i}{\rho}[(1+M_{i+1}\rho)^{\frac{r}{r_s L_i}}-1], M_i\cdot M_{i+1}\right). \label{OSTBC}
\end{equation}

\subsubsection{Deadline Missing Probability}

%Denote the amount of time the $n$th message spent at the $i$th hop
%(between the $i$th and $i+1$th node) message as $D_i^n$. The
%end-to-end delay is a sum of the per-hop delays: $D^n = \sum_{i=1}^{N-1}
%D_i^n$. Since the queues are stable, under the fixed ARQ protocol, the queueing process
%converges to a stationary process. The deadline missing probability is based on the stationary process, for which we have the following definition:
%\ben
%P_{deadline}(\{L_i\}|\rho, k) \triangleq P(\mbox{deadline missing}) =\lim_{n\rightarrow \infty} P(D^n>k). 
%\een

For a three-node network with half-duplex relay,   there is only one queue at the source that incurs the queueing delay. 
%The service time $S_n$ in this network is the \hl{duration} that the source transmitting to the relay and the relay transmitting to the
%destination, 
For given
$r$ and $\rho$, we can derive the stationary deadline missing probability using a martingale argument:
\begin{thm}\label{lemma1}
For a half-duplex three-node MIMO multihop network with Poisson arrival of rate $\lambda$ and ARQ rounds $L_1$ and $L_2$, the probability that the end-to-end delay exceeds  the deadline $k$ is given by
\begin{equation} P_{deadline}\lsb \{L_i\}|\rho, k\rsb
%\triangleq  \lim_{n\rightarrow\infty}P\lbb D_n > k \middle\vert \cap_{i=1}^{N-1}\left\{ l_n^i \leq L_i \right\} \rbb
= \frac{\mu(L_1)+\mu(L_2)}{\lambda}e^{-k \lsb \frac{1}{\mu(L_1)+\mu(L_2)} - \frac{1}{\lambda}
\rsb}. \label{three-hop-dd}
\end{equation} 
\label{lemma_stat_dist}
\end{thm}
\vspace{-0.4in}
\noindent Proof: See Appendix \ref{app:lemma_stat_dist}.

For general multihop networks with any number of half-duplex relays, the analysis for (\ref{deadline_finiteSNR}) is more involved. Due to half-duplexing the neighboring links cannot operate
simultaneously, and this effect is not captured in the standard queueing network analysis. Here we adapt the proof in \cite{Ganesh1998}, which uses large deviation techniques, to derive the following theorem for the exponent of the deadline missing probability in half-duplex relay networks:
\begin{thm}\label{SojournTimeHalfDuplex}
For a half-duplex $N$-node MIMO multihop network, with Poisson arrival of rate $\lambda$ and ARQ rounds $L_i$'s, the probability that the end-to-end delay exceeds the deadline $k$ is given by
\begin{eqnarray}
\lim_{k\rightarrow \infty} \lim_{n \rightarrow \infty}\frac{1}{k}
P\lbb D_n > k \middle\vert \cap_{i=1}^{N-1}\left\{ t_n^i \leq L_i \right\} \rbb
= -\theta^*,
\end{eqnarray}
where $\theta^* = \min_{1\leq i\leq N-2}\theta_i$, and $\theta_i = 1/[\mu(L_i)+\mu(L_{i+1})]-1/\lambda$, $i = 1, \cdots, N-2$.
\label{thm_half_duplex}
\end{thm}
\noindent Proof: See Appendix \ref{app:SojournTimeHalfDuplex}.

This theorem again demonstrates that the performance of the $N$-node multihop network with a half-duplex relay (here the performance metric is in terms of the deadline-missing probability exponent) is determined by the smallest exponent of each three-node sub-network. By Theorem \ref{thm_half_duplex}, for {f}inite $k$ we can approximate  (\ref{deadline_finiteSNR}) as
\begin{equation}
 P_{deadline}(\{L_i\}|\rho, k)\approx e^{-k\theta^*}. \label{more-hop-dd}
\end{equation}
Also note that in the special case with $N = 3$ nodes, for {f}inite $k$
\begin{equation}
 P_{deadline}(\{L_i\}|\rho, k)\approx e^{-k \left(\frac{1}{\mu(L_1)+\mu(L_2)}-\frac{1}{\lambda}\right)}, 
\end{equation}
which is identical with (\ref{three-hop-dd}) up to a multiplicative constant $[\mu(L_1)+\mu(L_2)]/\lambda$. This constant is typically not identi{f}iable by large deviation techniques such as the one used in Theorem \ref{thm_half_duplex}.

\subsubsection{Mean Service Time Calculation}

The above analysis requires $\mu(L_i)$, which we will derive in this section. For a given $t$ and message $n$, the cumulative distribution function (CDF) of $t_n^i$ is given by Ê
\begin{equation}
P\lbb t_n^i \leq t\rbb = 
P\lbb t C_i(\vect{H}_i) \geq  r\log_2(1+M_{i+1}\rho)\rbb = 1 - P_{out, i}(t|\rho),\label{CDF}
\end{equation}
where $P_{out, i}(t|\rho)$ is given in (\ref{per-hop-out}) (or a term in the summation in (\ref{OSTBC}) for OSTBC). Differentiating (\ref{CDF}) gives the desired probability density function (PDF):
\begin{eqnarray}
P\lbb t_n^i = t\rbb = \frac{M_i r}{\rho r_s t^2(M_i\cdot
M_{i+1}-1)!}f_i^{M_i\cdot
M_{i+1} - 1} e^{-f_i} (1+M_{i+1}\rho)^{\frac{r/t}{r_s}}\log_2(1+M_{i+1}\rho),
\end{eqnarray}
where $f_ i \triangleq \frac{M_i}{\rho}\left[(1+M_{i+1} \rho)^{\frac{r/t}{r_s}}-1\right]$. Using this we have
\begin{equation}
\mu(L_i) = \int_{t=1}^{L_i} t P\lbb t_n^i = t\rbb dt + L_i \int_{t = L_i}^\infty P\lbb t_n^i = t\rbb dt. \label{39}
\end{equation}
%The second term in (\ref{39}) is due to messages having an ARQ outage.

%For discrete $\bar{L}_i$ the mean service time can be defined similarly. 

\subsection{Optimal {F}ixed ARQ Design at {F}inite SNR}

Based on the above analysis, we formulate the optimal {f}ixed ARQ design in the {f}inite SNR regime as an optimization problem that allocates the total ARQ window size among hops to minimize the probability of message error subject to the queue stability and the end-to-end delay constraint $k$:
\begin{equation}
\begin{split}
\min_{\{L_i\}} & \quad P_{out}(\{L_i\}|\rho) + [1-P_{out}(\{L_i\}|\rho)]P_{deadline}(\{L_i\}|\rho, k) \\
\mbox{subject to} & \quad 1\leq \mu(L_i) \leq \lambda, \quad \sum_{i=1}^{N-1} L_i \leq k.
\end{split} \label{opt_formulation}
\end{equation}
The terms in (\ref{opt_formulation}) are given by (\ref{finite_SNR_pout}), (\ref{OSTBC}), (\ref{three-hop-dd}), and (\ref{more-hop-dd}). This optimization problem can be solved numerically. In particular, for a three-hop network with OSTBC, (\ref{opt_formulation}) becomes
\begin{eqnarray}
\min_{\{L_i\}} && \quad \sum_{i=1}^{2}\frac{1}{(M_i M_{i+1}-1)!}\gamma\left(\frac{M_i}{\rho}[(1+\rho)^{\frac{r}{r_s L_i}}-1], M_i M_{i+1}\right) + 
\frac{\mu(L_1)+\mu(L_2)}{\lambda}e^{-k \lsb \frac{1}{\mu(L_1)+\mu(L_2)} - \frac{1}{\lambda}
\rsb} \nonumber\\
\mbox{subject to} && \quad 1\leq \mu(L_i) \leq \lambda, \quad \sum_{i=1}^{N-1} L_i \leq k.
\label{opt_formulation_3hop}
\end{eqnarray}
%Note that in the formulation of (\ref{opt_formulation}) we no longer need an upper bound $L$ on the sum of ARQ window sizes over all hops. This upper bound is chosen heuristically in the asymptotic DMDT in Section III; in the current formulation it is taken care of by the end-to-end delay constraint $k$. 

As we demonstrate in the following examples,  the information outage probability is
decreasing in $L_i$, and the deadline missing probability is increasing in $L_i$. Hence the optimal ARQ
window size allocation $L_i$ at each node $i$ should trade off these
two terms. Moreover, the optimal ARQ window size allocation should equalize the
performance of each link.

\subsection{Numerical Example}

\begin{figure}[h]
    \begin{center}
        \includegraphics[width=2.5in]{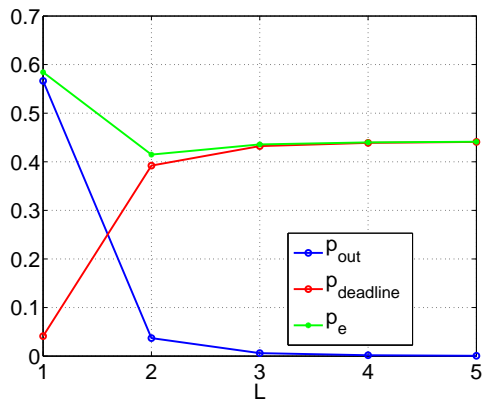}
           \end{center}
    \caption{The information outage probability and the deadline missing probability for a (2, 2) point-to-point MIMO system, with $r_s = 1$, $r = 1$, and SNR $\rho = 3$ dB, $k = 5$ blocks and $\lambda = 2$ blocks. The minimum probability of message error is $0.4147$ and is achieved by $L = 2$.}
    \label{Fig:tradeoff_finite_snr}
\end{figure}

%To illustrate the tradeoff between probability of information outage and the probability of deadline missing when assigning ARQ window sizes, 
We {f}irst consider a point-to-point (2, 2) MIMO system  at the source and the destination. Assume $\rho = 3$ dB and $\lambda = 2$ blocks. An OSTBC with $r_s = 1$ is used, for which the information outage probability is given in (\ref{OSTBC}). The deadline constraint is $k = 5$ blocks. For $r = 1$, the information outage probability  (\ref{out_finiteSNR}) and the deadline missing probability (\ref{deadline_finiteSNR}) are shown in {F}ig. \ref{Fig:tradeoff_finite_snr}. Note that (\ref{out_finiteSNR}) decreases, while (\ref{deadline_finiteSNR}) increases with the ARQ window size. 
% The minimum probability of message error is 0.4147 and is achieved by $L = 2$.  

\begin{figure}[h]
    \begin{center}
        \includegraphics[width=3in]{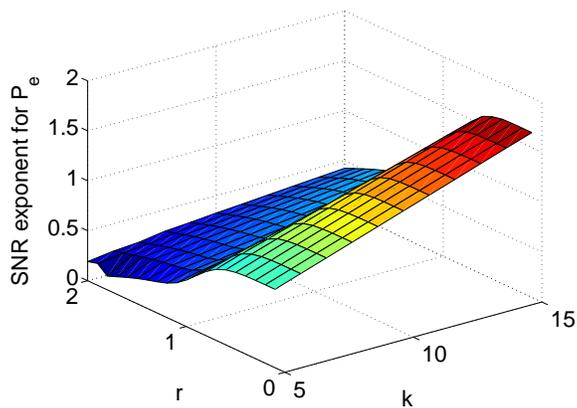}
    \end{center}
    \caption{The probability of message error for  the optimal {f}ixed ARQ protocol in a (4, 1, 3) multihop network as a function of multiplexing gain $r$ and delay constraint $k$; $r_s = 1$, SNR $\rho=20$ dB and $\lambda = 10$ blocks.}
\label{Fig:finite_snr_surface}
\end{figure}

Next we consider the (4, 1, 3) MIMO multihop network. Assume $\rho = 20$ dB and $\lambda = 10$ blocks. An OSTBC with $r_s = 1$ is used. The optimal {f}ixed ARQ protocol is obtained by solving (\ref{opt_formulation_3hop}).  For $r = 1$ and a deadline constraint of $k = 5$ blocks, the optimal {f}ixed ARQ has $L_1 = 2$ and $L_2 = 3$, and the optimal probability of message error is 0.1057. For $r = 1$ and $k = 10$ blocks, the optimal {f}ixed ARQ has $L_1 = 4$ and $L_2 = 6$, and the optimal probability of message error is 0.0355. For all $r$ and $k$, the probability of message error is plotted as a surface  in {F}ig. \ref{Fig:finite_snr_surface}. This surface is the DMDT for the three-node network in the {f}inite SNR regime, which has an interesting similarity to the high SNR asymptotic DMDT surface in {F}ig. \ref{Fig:dmt_413_L_slice}, since indeed the high SNR DMDT represents the SNR exponent of the {f}inite SNR DMDT. 

\section{Conclusions}\label{sec:conclusion}

We have analyzed the asymptotic diversity-multiplexing-delay tradeoff (DMDT)
for the $N$-node MIMO multihop relay network with ARQ, under both long-term and the short-term static channel assumptions.  We show that the asymptotic DMDT can be cast into an optimization problem that can be solved numerically in general, and closed-form asymptotic DMDT expressions are obtained in some special cases. We also proposed the VBL ARQ protocol which adapts the ARQ window size among hops dynamically and proved
that it achieves the optimal DMDT under both
channel assumptions. We also show that the DMDT for general multihop networks with multiple half-duplex relays can be found by decomposing the network into three-node sub-networks such that each sub-network consists of three neighboring nodes and its corresponding two hops. The DMDT of the relay network is determined by the minimum of the DMDTs of the three-node sub-networks. We have also shown that the DMDT of the three-node subnetwork is determined by its weakest link. Hence, the optimal ARQ
should equalize the link performances by properly allocating the per-hop ARQ window sizes dynamically.

We then studied the DMDT in the {f}inite SNR regime for {f}ixed ARQ protocols.  We introduced an end-to-end delay constraint such that a message is dropped
once its delay exceeds the delay constraint. Since in the {f}inite SNR regime retransmission is not a rare event, we incorporated the queueing delay into the system model, and modeled the system as a queueing network. The {f}inite SNR DMDT is characterized by the probability of message error, which consists of the information outage probability and the deadline missing probability. While the information outage probability can be found through {f}inite SNR DMDT analysis, we have also found the exponent for the deadline missing probability. Our result demonstrates that the performance of a multihop network with half-duplex relays in the {f}inite SNR regime is also determined by the performance of the weakest three-node sub-network. It has been shown that, based on these analyses, the optimal {f}ixed ARQ window size allocation can be solved numerically as an optimization problem, which should balance the per-hop diversity performance and  avoid a long per-hop delay.

The dif{f}iculty in
merging the network layer analysis with the physical layer information
theoretic results stems from the bursty nature of the source and the
end-to-end delays. By modeling the multihop relay network with ARQ
as a queuing network, we have tried to answer a
question posed at the end of \cite{HollidayGoldsmithPoor2008}: how to couple the fundamental performance limits of general multihop networks with queueing delay. Our work provides a step towards bridging the gap between network theory and information theory. Future work includes  developing an optimal dynamic ARQ protocol that can adapt to the channel state and the message arrival rate. The problem can be formulated as a dynamic programming problem or analyzed using a heavy traf{f}ic approximation.

\appendices

\section{Proof of Theorem \ref{thm_static_ARQ}}\label{app:thm_static_ARQ}

With {f}ixed-ARQ protocol and half-duplex relays, the system is in outage if any hop is in outage. The probability of message error $P_e(\rho)$,  using the decoding time de{f}inition in (\ref{stop_3}), can be written as:
\begin{eqnarray}
P_e(\rho) && \doteq P\lbb  \lbb t_1 > L_1 \rbb \cup  \lbb t_2 > L_2 \rbb  \rbb \label{app:eqn1}\\
&&=  P\lbb t_1 > L_1 \rbb + P\lbb t_2 > L_2 \rbb \label{app:eqn2} \\
&& \doteq  P \lbb L_1 S_1(\vect{\alpha}_1)< r\rbb
+  P \lbb L_2 S_2(\vect{\alpha}_2)< r\rbb\\
&&  \doteq \sum_{i=1}^{2}\rho^{-d^{(M_i, M_{i+1})}\lsb \frac{r}{L_i}\rsb} \label{app:eqn3}\\
&& \doteq \rho^{-\min_{i=1, 2} d^{(M_i, M_{i+1})}\lsb \frac{r}{L_i}\rsb},\label{d_multi_ARQ}
\end{eqnarray}
where  (\ref{app:eqn2}) is due to the independence of each link, and (\ref{app:eqn3}) follows from the method used in
\cite{ZhengTse03diversity},
%\cite{DemboZeitouni1992} ,
%as we did in defining (\ref{stop_3}), 
and the fact that 
\begin{equation}
P\lbb t_i > L_i \rbb = P\lbb \sum_{l=1}^{L_i} S_i(\vect{\alpha}_i^l) < r \rbb,
\end{equation}
since $S_i(\vect{\alpha}_i^l) \geq 0$ and $S_i(\vect{\alpha}_i^l) = S_i(\vect{\alpha}_i)$ for the long-term static channel.  The last equality follows since when SNR is high, the dominating
term is the one with the smaller SNR exponent. Using (\ref{d_multi_ARQ}) and the de{f}inition of diversity in (\ref{def_diversity})  we obtain the DMDT stated in Theorem
\ref{thm_static_ARQ}.

\section{Proof of Theorem \ref{thm_block_DDF}}\label{app:thm_block_DDF}
For the FBL ARQ protocol with two hops, the probability of
message error is given by
\begin{eqnarray}
P_{e}(\rho) \doteq P\lbb t_1 + t_2 > L \rbb = \sum_{k = 1}^{L-1}
P\lbb t_1 = k\rbb P\lbb t_2 > L - k\rbb + P\lbb t_1 \geq L \rbb.
\label{stop_formula}
\end{eqnarray}
%All the following proofs rely on the idea of ``stopping time
%formula'' (\ref{stop_formula}).
%
In the long-term static channel model we have
\begin{equation}
\begin{split}
P\lbb t_1 = k\rbb &= P \lbb (k-1)C_1 (\vect{H}_1) < r\log \rho \leq k C_1 (\vect{H}_1)\rbb  \\
& =  P \lbb C_1(\vect{H}_1) < \frac{r}{k-1}\log
\rho\rbb - P \lbb
C_1(\vect{H}_1) \leq \frac{r}{k}\log \rho\rbb  \\
&\doteq \rho^{-d^{(M_1, M_2)}\lsb \frac{r}{k-1}\rsb} - \rho^{-d^{(M_1, M_2)}\lsb
\frac{r}{k}\rsb}   \doteq  \rho^{-d^{(M_1, M_2)}\lsb
\frac{r}{k-1}\rsb},
\end{split}
\label{Pk}
\end{equation}
which follows from  the fact that $d^{(M_1, M_2)}(r)$ is monotone decreasing, i.e., $d^{(M_1, M_2)}\lsb
\frac{r}{k-1}\rsb \leq d^{(M_1, M_2)}\lsb \frac{r}{k}\rsb$. If we plug (\ref{Pk})
into (\ref{stop_formula}) we get
\begin{eqnarray}
P_{e}(\rho) &&\doteq P\lbb C_1(\vect{H}_1) \geq r\log\rho \rbb P
\lbb
C_2(\vect{H}_2) \leq \frac{r\log\rho}{L-1}\rbb \nonumber \\
&& + \sum_{k = 2}^{L-1} P \lbb(L-k)C_2(\vect{H}_2) < r\log \rho
\rbb P(t_1
= k) + P\lbb C_1(\vect{H}_1) <  \frac{r\log\rho}{L-1} \rbb \label{p_out_block_1}\\
&&\doteq \rho^{- \min_{k=2,\cdots,L-1}\lbb d^{(M_1, M_2)}\lsb
\frac{r}{L-1}\rsb, d^{(M_1, M_2)}\lsb \frac{r}{k-1}\rsb +d^{(M_2, M_3)}\lsb
\frac{r}{L-k}\rsb, d^{(M_2, M_3)}\lsb\frac{r}{L-1}\rsb\rbb}\nonumber\\
&& = \rho^{- \min_{ l_1+ l_2 = L-1, l_1 = \{0, 1, \cdots, L-1\}}, \lbb  d^{(M_1, M_2)}\lsb \frac{r}{l_1}\rsb +d^{(M_2, M_3)}\lsb
\frac{r}{l_2}\rsb\rbb},\label{outage_fix}
\end{eqnarray}
where we have used the fact that $d^{(M_i, M_{i+1})}(\infty) = 0$. From the de{f}inition of diversity in (\ref{def_diversity}) the DMDT in Theorem \ref{thm_block_DDF} follows. 
%\begin{flushright}$\blacksquare$\end{flushright}

\section{Proof of Theorem \ref{thm_DDF}}\label{app:thm_DDF} 

The decoding time of VBL ARQ is real, which differs from FBL ARQ. 
Since the long-term static channel has constant state, we can write the decoding
time as $t_i = \frac{r\log\rho}{C_i(\vect{H}_i)}.$ Hence:
\begin{equation}
\begin{split}
P_e(\rho) & \doteq   P\lbb t_1 + t_2 > L \rbb \\
&\doteq P\lbb (L - t_1)C_2(\vect{H}_2) < r\log \rho < L C_1(\vect{H}_1) \rbb  \\
& \doteq   P\lbb L
< \frac{r}{S_1(\vect{\alpha}_1)} + \frac{r}{S_2(\vect{\alpha}_2)} \rbb,
\end{split}\label{stopping_time_lt}
\end{equation}
and $d_{VBL}^{(M_1, M_2, M_3)}(r, L) =  \inf_{\lbb\alpha_{i,j}\rbb \in
\mathcal{O}} h(\lbb\alpha_{i,j}\rbb)$,
where $\mathcal{O}$ is de{f}ined in (\ref{feas_VBL2}).

To prove that the DMDT of VBL ARQ is the optimal DMDT in an $N$-node network, we {f}irst provide an upper
bound on the DMDT, and show
that the DMDT of the VBL ARQ protocol achieves this upper bound. Our proof is for the short-term static channel in a three-node network, as stated in Theorem \ref{thm_st_fractional}. A similar (and simpler) proof can be written for the long-term static channel in a three-node network as stated in Theorem \ref{thm_DDF}. 
Assume that the source transmits for $k T$ channel
uses ($kT < L$) and the relay transmits in the remaining $L - k T $ channel uses. Here $k$ depends on the channel states and the multiplexing gain $r$.
From the cut-set bound on the multihop network channel capacity, the instantaneous capacity of the MIMO
ARQ channel is given by
\begin{equation}
\begin{split}
&\min \lbb 
\max_{P_{\vect{X}_{1, l }, l = 1, \cdots, \lfloor k \rfloor +1}} \lmb
 \sum_{l =1}^{ \lfloor k \rfloor}
I(\vect{X}_{1,l}; \vect{Y}_{1,l}| \vect{H}_{1,l}) +(k - \lfloor k \rfloor)
I(\vect{X}_{1,k+1}; \vect{Y}_{1,k+1}| \vect{H}_{1,k+1})
\right.\rmb,  \\
&~~~\left. \max_{P_{\vect{X}_{2, l}, l= 1, \cdots, L-\lfloor k \rfloor-1}} \lmb \sum_{l = 1}^{L-\lfloor k \rfloor-1} I(\vect{X}_{2,l}; \vect{Y}_{2,l}|
\vect{H}_{2,l}) + (1-k+\lfloor k \rfloor)I(\vect{X}_{2,L-k}; \vect{Y}_{2,L-k}|
\vect{H}_{2,L-k}) \rmb \rbb.
\end{split}\nonumber
\end{equation}
Since the capacity is maximized with Gaussian inputs, and linear
scaling of the power constraint does not affect the high SNR
analysis, the capacity $C$ is upper bounded by
\begin{eqnarray}
C \leq \min \lbb \sum_{l =1}^{ \lfloor k \rfloor} C_1(\vect{H}_{1,l}) + (k-\lfloor k \rfloor)
C_1(\vect{H}_{1,k+1}), \sum_{l = 1}^{L-\lfloor k \rfloor-1} C_2(\vect{H}_{2,l}) +
 (1-k+\lfloor k \rfloor)C_2(\vect{H}_{1,L-k}) \rbb.
\end{eqnarray}
For any ARQ we can {f}ind a $k^*< L$  such that $
\sum_{l =1}^{ k^*} C_1(\vect{H}_{1,l}) + (k^* -  \lfloor k^* \rfloor)
C_1(\vect{H}_{1,k^*+1}) = r\log\rho $. This means $k^* \doteq t_1$. 
With this $k^*$, the probability of message error is lower bounded by
\begin{equation}
\begin{split}
 P_{e}(\rho)   
&\geq P\lbb r\log \rho >  \sum_{l =
1}^{L-\lfloor k^*\rfloor -1} C_2(\vect{H}_{2,l}) +
(1- k^* + \lfloor k^*\rfloor)C_2(\vect{H}_{1,L-\lfloor k^* \rfloor})  \rbb \\
&\doteq P\lbb 
\lbb r >   \sum_{l = 1}^{L-\lfloor t_1\rfloor -1} S_2(\vect{\alpha}_{2}^l) +
(1-t_1 + \lfloor t_1\rfloor)S_2(\vect{\alpha}_{2}^{L-\lfloor t_1\rfloor}) \rbb\cap \tilde{\mathcal{G}} \rbb \\
&= P\lbb \lbb t_2 > L - t_1 \rbb \cap \tilde{\mathcal{G}}  \rbb =P\lbb \lbb t_1 + t_2 > L \rbb \cap \tilde{\mathcal{G}}\rbb,
\end{split}
\end{equation}
where $\tilde{\mathcal{G}} = \lbb (\vect{\alpha_1},\cdots, \vect{\alpha}_{N-1})\in\mathbb{R}^{M_1^* \times
L}\times\cdots \times\mathbb{R}^{M_{N-1}^*\times L}: 
 \alpha_{i,
1}^l\geq \cdots \geq \alpha_{i, M_i^*}^l \geq 0, \forall i,
l
\rbb$.
Hence, the diversity gain of any ARQ $d^{(M_1, M_2, M_3)}(r, L)$ of a three-node network  is upper bounded by
\begin{eqnarray}
d^{(M_1, M_2, M_3)}(r, L) \dot{\leq} \inf_{\alpha_{i,j}^l \in\mathcal{G}_2}
\tilde{h}(\alpha_{i,j}^l), \label{app_3}
\end{eqnarray}
with $\mathcal{G}_2 \triangleq \lbb t_1 + t_2 > L \rbb \cap \tilde{\mathcal{G}}$, which is the same as the set $\mathcal{G}$ in (\ref{VBL_region}), the DMDT expression for 
VBL ARQ in the short-term static channel. This shows that the DMDT upper bound is achieved by the VBL ARQ in the short-term static channel. This completes our proof.

\section{Proof of Theorem \ref{thm_st}}\label{app:thm_st}

In the short-term
static channel, for the FBL ARQ protocol with two hops
(\ref{Pk}) becomes
\begin{equation}
\begin{split}
P\lbb t_1 = k\rbb &= P \lbb \sum_{l = 1}^{k-1}C_1 (\vect{H}_{1,l}) <
r\log \rho \rbb - P \lbb \sum_{l = 1}^{k}C_1 (\vect{H}_{1,l}) <
r\log \rho \rbb  \\
&\doteq \lsb P \lbb S_1 (\vect{\alpha}_1^1) < \frac{r\log \rho}{k-1}
\rbb \rsb^{k-1} - \lsb P \lbb S_1 (\vect{\alpha}_1^1) < \frac{r\log
\rho}{k} \rbb \rsb^{k} \doteq \rho^{-(k-1) d^{(M_1, M_2)}\lsb
\frac{r}{k-1}\rsb}.
\end{split}\nonumber
\end{equation}
Hence, the probability of message error can be written as
\begin{eqnarray}
P_{e}(\rho) &\doteq& P\lbb C_1(\vect{H}_{1,1}) \geq r\log\rho \rbb P
\lbb
\sum_{l = 2}^{L} C_2(\vect{H}_{2,l}) \leq r\log\rho \rbb \nonumber \\
&+& \sum_{k = 2}^{L-k} P \lbb\sum_{l = k+1}^{L}C_2(\vect{H}_{2,l})
< r\log \rho \rbb P(t_1
= k) + P\lbb \sum_{l =1}^{L} C_1(\vect{H}_{1, l}) \geq r\log\rho \rbb  \nonumber \\
&\doteq& \rho^{- \min_{k=2,\cdots,L-1}\lbb (L-1) d^{(M_1, M_2)}\lsb
\frac{r}{L-1}\rsb, (k-1)d^{(M_1, M_2)}\lsb \frac{r}{k-1}\rsb + (L-k)d^{(M_2, M_3)}\lsb
\frac{r}{L-k}\rsb, (L-1)d^{(M_2, M_3)}\lsb\frac{r}{L-1}\rsb\rbb}. \nonumber
%\label{p_out_st}
\end{eqnarray}
By the de{f}inition of diversity, the DMDT in Theorem \ref{thm_st} follows.

\section{Proof of Theorem \ref{thm_st_fractional}}\label{app:thm_st_fractional} For a three-node network, we can break down the
information outage event as a disjoint union of events that outage
happens at the $i$th hop: 
\begin{equation}
P_{e}(\rho) \doteq
\rho^{-\inf_{\lbb\alpha_{i,j}^l\rbb \in \cup_{k=1}^2
\mathcal{G}_k} \tilde{h}\left(\lbb\alpha_{i,j}^l\rbb\right)}, 
\end{equation}
where
$\mathcal{G}_k \triangleq \lbb \sum_{i=1}^{k}t_i
>L\rbb$. Due to nonnegativity of $t_i$,
$\mathcal{G}_1\subset\mathcal{G}_2$. Hence,
the minimization should be over $\mathcal{G}_2$. Adding the
ordering requirement on elements of $\lbb\alpha_{i,j}^l\rbb$, we have
Theorem \ref{thm_st_fractional}.
%\begin{flushright}$\blacksquare$\end{flushright}

\section{Proof of Theorem \ref{thm_n_VBL}}\label{app:thm_n_VBL}

\subsection{Upper bound}\label{app:upper_bound}

We will {f}irst prove an upper bound for any ARQ protocol in an $N$-hop network by considering a genie-aided scheme. 
For each $i = 1, \cdots, N - 2$, consider the two consecutive hops from node $i$ to node $(i +1)$ and then from node $(i+1)$ to node $(i + 2)$. Assume a genie aided scheme where the messages are provided to node $i$, and the output of node $(i+2)$ is forwarded to the terminal node $N$.  The maximum  number of ARQ rounds that can be spent on this two-hop is $L$.  The DMDT of this genie aided setup for any $i$, is an upper bound on the DMDT of the $(M_1, \cdots, M_{N})$ system. The optimal DMDT of the $(M_i, M_{i+1}, M_{i+2})$ system with $L$ ARQ rounds is characterized in Theorem \ref{thm_DDF}. Hence, we have
\begin{eqnarray}
d^{(M_1, \cdots, M_{N})}(r, L) 
&\leq \min_{i = 1, \cdots, N-2} d^{(M_i, M_{i+1}, M_{i+2})}_{VBL} (r, L),\label{eqn1}
\end{eqnarray}
where $d^{(M_1, \cdots, M_{N})}(r, L)$ is the DMDT of any ARQ protocol for an $N$-hop network. 

\subsection{The DMDT of VBL ARQ}\label{achievable_VBL}

To be able to exploit the multi-hop diversity in the network, we use the following rate and ARQ round allocation scheme. {F}irst we split the original message of rate $r\log\rho$ into $N/2$ lower rate messages  each having a rate of $(r\log\rho)/(N/2)$ when $N$ is even (we split into $(N-1)/2$ lower rate messages when $N$ is odd). We pump these pieces of the original message into the network sequentially, and in equilibrium, they are transmitted simultaneously by adjacent pairs of nodes.

Moreover, we require the number of blocks allowed for any two-hop transmission, from node $i$ to node $(i+1)$ and then to node $(i+2)$, for all $i = 1, \cdots, N-2$,  to be $\bar{L} = L/(N/2)$ when $N$ is even (or $\bar{L} = L/[(N-1)/2]$ when $N$ is odd). This is equivalent to requiring the total number of blocks that each node $i$, $i = 2, \cdots, N$, spends for  listening and transmitting each piece of a message to be $\bar{L}$. Note that with this constraint, the end-to-end total number of ARQ rounds used for  transmitting each piece of the original message is upper bounded by $\bar{L}\times N/2 = L$ when $N$ is even (or equals $\bar{L}\times (N-1)/2  = L$ when $N$ is odd). Hence, this scheme satis{f}ies the constraint on the end-to-end total number of ARQ rounds.

It is easy to see that the number of simultaneous transmission pairs we can have in an $N$ node network is $N/2$ when $N$ is even, and $(N-1)/2$ or $(N+1)/2$ when $N$ is odd.

At the destination, all pieces of a message are combined to decode the original message. From the above analysis, the last piece of these low rate messages is received after at most $L$ blocks, and the rate of the combined data is $r\log\rho/(N/2)\times(N/2) = r\log \rho$ when $N$ is even (similarly for odd $N$), which equals the original data rate. Hence this low rate message simultaneous transmission scheme meets both the data rate and end-to-end ARQ window size constraints.

Now we study the outage probability $P_{out}(r)$ of this scheme. De{f}ine an outage event for any three-node sub-network consisting of nodes $i$, $(i+1)$, and $(i+2)$, for $N$ even, as:
\begin{equation}
\begin{split}
P_{out}^i(r, L)& \triangleq P\left\{\frac{r/(N/2)\log\rho}{C_i(\vect{H}_i)} +  \frac{r/(N/2)\log\rho}{C_{i+1}(\vect{H}_{i+1})} > \frac{L}{(N/2)}\right\} \\
& = P\left\{\frac{r\log\rho}{C_i(\vect{H}_i)} +  \frac{r\log\rho}{C_{i+1}(\vect{H}_{i+1})} > L \right\}, 
\end{split}\label{eqn_even}
\end{equation}
and for $N$ odd, similarly, as
\begin{equation}
\begin{split}
P_{out}^i(r, L)& \triangleq P\left\{\frac{r/[(N-1)/2]\log\rho}{C_i(\vect{H}_i)} +  \frac{r/[(N-1)/2]\log\rho}{C_{i+1}(\vect{H}_{i+1})} > \frac{L}{[(N-1)/2]}\right\} \\
& = P\left\{\frac{r\log\rho}{C_i(\vect{H}_i)} +  \frac{r\log\rho}{C_{i+1}(\vect{H}_{i+1})} > L \right\}.
\end{split}\label{eqn_odd}
\end{equation}
Note that (\ref{eqn_even}) and (\ref{eqn_odd}) say that by using this scheme, regardless of whether $N$ is even or odd, the outage probability is as if we transmit the original message with data rate $r\log\rho$ over two hops with a total ARQ round constraint of $L$. From our earlier analysis for the VBL ARQ of a two-hop network, we have that as $\rho\rightarrow \infty$,
\begin{equation}
P_{out}^i(r, L) \doteq \rho^{-d_{VBL}^{(M_i, M_{i+1}, M_{i+2})}(r, L)}.
\end{equation}

The system is in outage if there is an outage over any of the consecutive two-hop links from the source to the destination. Using a union bound, we have
\begin{equation}
P_{out}(r, L) \leq \sum_{i=1}^{N-2} P_{out}^i (r, L).
\end{equation}
As SNR goes to in{f}inity, the right hand sum will be dominated by the slowest decaying term, which is the term with minimum $d_{VBL}^{(M_i, M_{i+1}, M_{i+2})}(r, L)$. Hence, the DMDT of this scheme is lower bounded by
\begin{equation}
d^{(M_1, \cdots, M_{N})}(r, L) \geq \min_{i = 1, \cdots, N-2} d_{VBL}^{(M_i, M_{i+1}, M_{i+2})}(r, L).
\end{equation}

Together with the upper bound in (\ref{eqn1}), this shows that the presented scheme with the VBL ARQ achieves the optimal DMDT of an $N$-hop network, and its DMDT is given by Theorem \ref{thm_n_VBL}.

\section{Proof of Theorem \ref{thm_n_fixed}}\label{app:thm_n_fixed}

The proof for the upper bound is identical to the one in Appendix \ref{app:upper_bound}. For the achievable DMDT of the {f}ixed ARQ, we consider the following scheme with deterministic number of ARQ rounds: a node has to wait for at least $L_i$ rounds over hop $i$ for each piece of message, and we allow simultaneous transmissions to employ multihop diversity. Using this scheme, in steady state, the destination will receive one piece of the message every $L_{\max}$ rounds (rather than $L$, if we do not employ multihop diversity).  Now we divide the message into pieces with lower rates $\frac{L_{\max}}{L}r\log\rho$. Using this scheme, overall we will still achieve a rate of $r\log\rho$ in the steady state by transmitting these lower rate pieces. The outage probability of this scheme provides an upper bound on that of the {f}ixed ARQ protocol:
\begin{equation}
P_{out}(r, L_1, \cdots, L_{N-1})  \leq \sum_{i=1}^{N-1} 
P\left\{\left\lceil\frac{\frac{L_{\max}}{L}r\log\rho}{C_i(\vect{H}_i)}\right\rceil
> L_i
\right\},
\end{equation}
where $\lceil x \rceil$ is the smallest integer larger than $x$.
As SNR goes to in{f}inity, the right hand sum will be dominated by the slowest decaying term, which is the term with minimum $d^{(M_i, M_{i+1})}\left(\frac{L_{\max}}{L}\cdot\frac{r}{L_i} \right)
$, and, hence,
\begin{equation}
\begin{split}
d_{F}^{(M_1, \cdots, M_N)}(r, L_1, \cdots, L_{N-1}) &\geq 
\min_{i=1, \cdots, N - 1} d^{(M_i, M_{i+1})}\left(\frac{L_{\max}}{L}\cdot\frac{r}{L_i} \right)
\\
&=\min_{i = 1, \cdots, N-2} d_{F}^{(M_i, M_{i+1}, M_{i+2})}\left(\frac{L_{\max}}{L}r, L_i, L_{i+1}
\;\middle\vert\;
L_i + L_{i+1} \leq L_{\max}\right),
\end{split}
\end{equation}
which completes our proof. 

\section{Proof of Theorem \ref{thm_n_FBL}}\label{app:thm_n_FBL}
The proof for the upper bound is identical to Appendix \ref{app:upper_bound}. For the achievable DMDT of the {f}ixed ARQ, again consider the same rate-splitting scheme in Appendix \ref{achievable_VBL}. The difference here is that the number of ARQ rounds used is rounded up to be integer. For $N$ even, the outage probability can be written as
\begin{equation}
\begin{split}
P_{out}^{i}(r, \bar{L}) 
&= P\left\{\lc \frac{r/(N/2)\log(\rho)}{C_i(\vect{H}_i)} \rc +  \lc\frac{r/(N/2)\log(\rho)}{C_{i+1}(\vect{H}_{i+1})}\rc > \bar{L} \right\} \\
& < P\left\{\frac{r/(N/2)\log(\rho)}{C_i(\vect{H}_i)} + 1 +
 \frac{r/(N/2)\log(\rho)}{C_{i+1}(\vect{H}_{i+1})} + 1 > \bar{L}  \right\} \\
& = P\left\{\frac{r\log(\rho)}{C_i(\vect{H}_i)} + \frac{r\log(\rho)}{C_{i+1}(\vect{H}_{i+1})} + N> L\right\} \\
&\doteq \rho^{-d_{VBL}^{(M_i, M_{i+1}, M_{i+2})}(r, L -N)}.
\end{split}
\end{equation}
Note that $L > N$ since we need at least $N$ hops.  
%
%Compare this with the outage probability of the FBL ARQ for a $(M_i, M_{i+1}, M_{i+2})$ networks, which is given by
%\begin{equation}
%\begin{split}
%P_{out}^{(M_i, M_{i+1}, M_{i+2})}(r, L)&= P\left\{\lc \frac{r\log(\rho)}{C_i(\vect{H}_i)} \rc + \lc\frac{r\log(\rho)}{C_{i+1}(\vect{H}_{i+1})}\rc > L\right\} \\
%& > P\left\{\frac{r\log(\rho)}{C_i(\vect{H}_i)}  + \frac{r\log(\rho)}{C_{i+1}(\vect{H}_{i+1})} > L\right\}
%\end{split}
%\end{equation}
%
The system is in outage if any three-node sub-network in outage. Using the union bound, again we have
\begin{equation}
P_{out}(r, L) \leq \sum_{i=1}^{N-2} P_{out}^{i}(r, \bar{L}) \leq \sum_{i=1}^{N-2} \rho^{-d_{VBL}^{(M_i, M_{i+1}, M_{i+2})}(r, L -N)}, 
\end{equation}
and
\begin{equation}
d_{FBL}^{(M_1, \cdots, M_N)}(r, L) \geq \min_{i = 1, \cdots, N-2} d_{VBL}^{(M_i, M_{i+1}, M_{i+2})}(r,  L - N).
\end{equation}

\section{Proof of Theorem \ref{lemma_stat_dist}}\label{app:lemma_stat_dist}

Theorem \ref{app:lemma_stat_dist} can be proved using Theorem 7.4.1 of
\cite{Ross1995}. The theorem views the random queueing delay of the $n$th message as a re{f}lected random walk. The deadline missing probability can be interpreted as a boundary hitting probability of the random walk, which can be obtained via a standard martingale argument which will not be repeated here. Note that the service time in a half-duplex two-hop network has a mean service time  of $\mu(L_1)+\mu(L_2)$ blocks (approximate the service time as exponentially distributed) when conditioned on the event $\cap_{i=1}^{N-1}\{t_n^i \leq L_i\}$. The mean message inter-arrival time is $\lambda$ blocks. Using the mean service time, the mean inter-arrival time, and the delay deadline constraint $k$ in  Theorem 7.4.1 of \cite{Ross1995}, we obtain the statement in Theorem \ref{lemma_stat_dist}.

\section{Proof of Theorem \ref{SojournTimeHalfDuplex}}\label{app:SojournTimeHalfDuplex}

The following proof is adapted from the proof in \cite{Ganesh1998}, where a conventional queue tandem is considered. The conventional queue tandem is equivalent to a full-duplex multihop network, where the transmission (service) of node $i$ for a message has to wait for the transmission of the previous message from node $i$ to node $(i+1)$. However, in our problem, we have a half-duplex scenario, in particular, the transmission (service) of node $i$ for a message has to wait for the transmission of the previous message over node $(i+1)$ to node $(i+2)$. The half-duplex scenario leads to a different and more complex queueing dynamic that we will study more precisely in the following. 

For node $i$,
$i = 1\cdots N-1$, $N\geq 3$, let the random variable $S_n^i$ denote the
service time of the $n$th message at node $i$, and $A_n^i$ be the
inter-arrival time of the $n$th message at node $i$. Due to the half-duplex constraint, there are $N-2$ queues at the source and node $i$, $i = 2, \cdots, N-2$.
After the completion of transmission of the previous message, the message will be transmitted from node $i$ to node $(i+1)$ and to node $(i+2)$, for $i = 1, \cdots, N-2$.  Because of this queueing dynamic, the
waiting time of the $n$th message at node $i$, $W_n^i$,
satis{f}ies the following form of Lindley's recursion (see \cite{Ganesh1998}):
\begin{equation}
W_n^i = (W_{n-1}^i + S_{n-1}^i + S_{n-1}^{i+1} - A_n^i)^+, \quad 
i = 1, \cdots, N-2, 
\label{rec_2}
\end{equation}
where $(x)^+ = \max(x, 0)$.  The total time a message spent for transmission from node $i$ to node $i+2$ is given by the waiting time plus its own transmission time
\begin{eqnarray}
D_n^i &= W_n^i + S_n^i + S_n^{i+1}, \quad i = 1, \cdots, N-2. \label{D_n_i}
\end{eqnarray}
Note there are overlaps in these transmission times $D_n^i$'s we de{f}ined above, so their sum provides an upper bound on the end-to-end delay of the $n$th message:
\begin{equation}
D_n \leq \sum_{i=1}^{N-2} D_n^i. \label{upper_bound}
\end{equation}

Next we will write $D_n^i$'s in (\ref{upper_bound}) explicitly using a non-recursive expression. The arrival process at node
$i$ is the departure process from node $(i-1)$, which
satis{f}ies the recursion \cite{Ganesh1998}:
\begin{eqnarray}
A_n^i = A_n^{i-1} + D_n^{i-1} - D_{n-1}^i,  \quad i = 1, \cdots, N-2,
\label{rec_1}
\end{eqnarray}
where $A_n^i$ is a Poisson process with mean interarrival time $\lambda$. A well-known result from queueing theory \cite{Ganesh1998} states the following: if the
arrival and service processes satisfy the stability condition, that is, the mean inter-arrival time  $\lambda$ is greater than the mean service time $\mu(L_i) + \mu(L_{i+1})$ at each of the queues $i = 1, \cdots, N-2$,
then Lindley's recursion (\ref{rec_2}) has the solution:
\begin{equation}
W_n^i  = \max_{j_i\leq n}(\sigma^{i}_{j_i, n-1} + \sigma^{i+1}_{j_i, n-1} - \tau^i_{j_i+1,n}),
\quad i = 1, \cdots N-2, \label{W_n_i}
\end{equation}
where the partial sums are de{f}ined as $\sigma_{l,p}^i = \sum_{k=l}^p S_k^i$ and $\tau_{l,p}^i = \sum_{k=l}^p A_k^i$. Hence, from (\ref{D_n_i}) and (\ref{W_n_i}), we have
\begin{eqnarray}
D_n^i =  \max_{j_i\leq n}(\sigma^{i}_{j_i, n} + \sigma^{i+1}_{j_i, n} - \tau^i_{j_i+1,n}), \quad i = 1, \cdots N-2.
\label{D}
\end{eqnarray}
One the other hand, from (\ref{rec_1}) we have 
\begin{equation}
\tau_{l,p}^i = \left\{
\begin{array}{lc}
\tau_{l,p}^{i-1} +
D_p^{i-1} - D_{l-1}^{i-1}, & l\leq p+1,\\
0, & \mbox{otherwise.}
\end{array}\right.\label{73}
\end{equation} Plugging
(\ref{73}) into  (\ref{D}), we have
\begin{eqnarray}
D_n^i = \max_{j_i\leq n}(\sigma^i_{j_i, n} + \sigma^{i+1}_{j_i, n}  -
\tau_{j_i+1,n}^{i-1} - D_n^{i-1} + D_{j_i}^{i-1}),\quad i = 2, \cdots, N-2.
\end{eqnarray}
Moving $D_n^{i-1}$ to the
left-hand-side, we obtain the recursion relation
\begin{eqnarray}
D_n^i + D_n^{i-1} = \max_{j_i\leq n}(\sigma_{j_i, n}^{i}+\sigma_{j_i, n}^{i+1} -\tau_{j_i+1,n}^{i-1} + D_{j_i}^{i-1}), \quad i=2,\cdots, N-2. \label{seq}
\end{eqnarray}
Now from (\ref{D}) we have $
D_{j_i}^{i-1} = \max_{j_{i-1}\leq
j_i}(\sigma^{i-1}_{j_{(i-1)},j_i} +\sigma^i_{j_{(i-1)},j_i}  -
\tau_{j_{(i-1)}+1, j_i}^{i-1})$. Plugging this into
(\ref{seq}) we have
\begin{eqnarray}
D_n^i + D_n^{i-1}
= \max_{j_{(i-1)}\leq j_i \leq n} (\sigma_{j_i, n}^{i} +\sigma_{j_i, n}^{i+1} 
+ \sigma^{i-1}_{j_{(i-1)},j_i} +\sigma^i_{j_{(i-1)},j_i}  - \tau_{j_{(i-1)}+1,
n}^{i-1}), \quad i = 2, \cdots, N-2.
\end{eqnarray}
Repeating this operation inductively by expanding $\tau_{j_{(i-1)}+1,
n}^{i-1}$, we obtain
\begin{eqnarray}
\sum_{i=1}^{N-2} D_n^i = \max_{j_1\leq \cdots \leq
j_{N-1}=n}\left[ \sum_{i=1}^{N-2}(\sigma^{i}_{j_i,j_{(i+1)}}+\sigma^{i+1}_{j_i,j_{(i+1)}} ) -\tau^1_{j_1+1, n} \right]. \label{recursion_1}
\end{eqnarray}

Note that in the non-recursive expression (\ref{recursion_1}), we split the interval $[1, n]$ by increasingly ordered integers $j_1, \cdots, j_{(N-1)} = n$, and the summations of random variables over these different intervals are mutually independent. This decomposition enables us to adopt a similar large deviation argument as in \cite{Ganesh1998}, to estimate the exponent $\theta^*$ in the form of $P(\sum_{i=1}^{N-2}D_n^i \geq k) \approx \exp(-\theta^* k)$ for large $k$. Following a similar argument  as in \cite{Ganesh1998} by {f}inding a condition under which the log-moment generating function is bounded for each independent sum-of-random-variables when $n\rightarrow \infty$, we can show that 
\begin{equation}
\lim_{k\rightarrow \infty}\frac{1}{k} P\left(\sum_{i=1}^{N-2} D_n^i \geq k\right)
 = -\min_{i=1, \cdots, N-2} \theta_i,
\end{equation}
where 
$\theta_i = \sup\{\theta > 0: \Lambda_T(-\theta) + \Lambda_{(S^i + S^{i+1})}(\theta)<0 \}$, $i = 1, \cdots, N-2$, and the log-moment-generating functions  for the inter-arrival time $\Lambda_T(\theta) = \log(1-\theta\lambda)^{-1}$, for the service time $\Lambda_{(S^i+ S^{i+1})}(\theta) = \log(1-\theta[\mu(L_i) + \mu(L_{i+1})])$ (if we approximate the sum service time to be exponentially distributed).
We can further solve that $\theta_i = (\mu(L_i) + \mu(L_{i+1}))^{-1} - \lambda^{-1}$, $i = 2, \cdots, N-2$. Because of (\ref{upper_bound}), $P(D_n \geq k) \leq P(\sum_{i=1}^{N-2} D_n^i \geq k)$, and hence the exponent for the deadline missing probability is bounded by $-\theta^* \leq -\min_{i=1, \cdots, N-2}\theta_i$. 

Now we prove the lower bound. Note that the end-to-end delay $D_n$ is greater than the delay in any three-node sub-network: $D_n \geq D_n^i$, $i = 1, \cdots, N-2$. Using a similar argument, we can show that the exponent for the probability that the delay in a three-node sub-network exceeds $k$ is given by $-\theta_i$. Hence, we have
\begin{equation}
-\theta^* = \lim_{k\rightarrow \infty}\frac{1}{k}\log P(D_n \geq k) \geq 
\lim_{k\rightarrow \infty}\frac{1}{k} \log P(D_n^i \geq k) =  -\theta_i.\label{81}
\end{equation}
Inequality (\ref{81}) still holds if we take the maximum over all $i$ on the right-hand-side:
\begin{equation}
-\theta^* \geq \max_{i=1, \cdots, N-1} -\theta_i = - \min_{i=1, \cdots, N-1} \theta_i.
\end{equation}
This completes our proof.

\bibliography{yao_proposal_2}

\section*{Acknowledgment}
The authors would like to thank J. Michael Harrison for his helpful suggestions and comments, which was of great help in our queuing analysis.

\end{document}